\documentclass[aps,prb,superscriptaddress,reprint,showpacs,amssymb,amsmath,floatfix,citeautoscript,longbibliography]{revtex4-2}

\usepackage{graphicx}
\usepackage{color}
\usepackage{amsmath}
\usepackage{float}
\usepackage{dsfont}
\usepackage{braket}
\usepackage{array}

\newcolumntype{C}[1]{>{\centering\let\newline\\\arraybackslash\hspace{0pt}}m{#1}}

\begin{document}

\title{Orbital selective coupling in CeRh$_3$B$_2$: co-existence of high Curie and high Kondo temperature}

\author{Andrea~Amorese}
    \altaffiliation{ASML Netherlands B.V., De Run 6501, 5504 DR, Veldhoven, The Netherlands}
		\affiliation{Institute of Physics II, University of Cologne, Z\"{u}lpicher Str. 77, 50937 Cologne, Germany}
    \affiliation{Max Planck Institute for Chemical Physics of Solids, N{\"o}thnitzer Str. 40, 01187 Dresden, Germany}
\author{Philipp~Hansmann}
    \affiliation{Department of Physics, Friedrich-Alexander-Universit\"at Erlangen-N\"urnberg, 91058, Erlangen, Germany}
\author{Andrea~Marino}
    \affiliation{Max Planck Institute for Chemical Physics of Solids, N{\"o}thnitzer Str. 40, 01187 Dresden, Germany}
\author{Peter K\"{o}rner}
  \altaffiliation{Subventa GmbH, Kurzes Gel{\"a}nd 14, 86156 Augsburg, Germany}
  \affiliation{Institute of Physics II, University of Cologne, Z\"{u}lpicher Str. 77, 50937 Cologne, Germany}
\author{Thomas Willers}
  \altaffiliation{KR{\"U}SS GmbH, Borsteler Chaussee 85, 22453 Hamburg, Germany}
 \affiliation{Institute of Physics II, University of Cologne, Z\"{u}lpicher Str. 77, 50937 Cologne, Germany}
\author{Andrew Walters}
    \affiliation{Diamond Light Source, Harwell Science \& Innovation Campus, Didcot, Oxfordshire OX11 0DE, United Kingdom}
\author{Ke-Jin Zhou}
    \affiliation{Diamond Light Source, Harwell Campus, Didcot, Oxfordshire OX11 0DE, United Kingdom}
\author{Kurt~Kummer}
  \affiliation{European Synchrotron Radiation Facility, 71 Avenue des Martyrs, CS40220, F-38043 Grenoble Cedex 9, France}
\author{Nicholas~B.~Brookes}
  \affiliation{European Synchrotron Radiation Facility, 71 Avenue des Martyrs, CS40220, F-38043 Grenoble Cedex 9, France}
\author{Hong-Ji~Lin}
 \affiliation{National Synchrotron Radiation Research Center (NSRRC), Hsinchu 30077, Taiwan}
\author{Chien-Te~Chen}
 \affiliation{National Synchrotron Radiation Research Center (NSRRC), Hsinchu 30077, Taiwan}
\author{Pascal~Lejay}
    \affiliation{Institut N\'{e}el, CNRS, BP166, 38042 Grenoble Cedex 9, France}
\author{Maurits~W.~Haverkort}
     \affiliation{Institute for Theoretical Physics, Heidelberg University, Philosophenweg 19, 69120 Heidelberg, Germany}
\author{Liu~Hao~Tjeng}
    \affiliation{Max Planck Institute for Chemical Physics of Solids, N{\"o}thnitzer Str. 40, 01187 Dresden, Germany}
\author{Andrea~Severing}
        \affiliation{Institute of Physics II, University of Cologne, Z\"{u}lpicher Str. 77, 50937 Cologne, Germany}
        \affiliation{Max Planck Institute for Chemical Physics of Solids, N{\"o}thnitzer Str. 40, 01187 Dresden, Germany}

\date{\today}

\begin{abstract}
We investigated the electronic structure of the enigmatic CeRh$_3$B$_2$ using resonant inelastic scattering and x-ray absorption spectroscopy in combination with \textit{ab-initio} density functional calculations. We find that the Rh\,4$d$ states are irrelevant for the high-temperature ferromagnetism and the Kondo effect. We also find that the Ce\,4$f$ crystal-field strength is too small to explain the strong reduction of the Ce magnetic moment. The data revealed instead the presence of two different active Ce\,4$f$ orbitals, with each coupling selectively to different bands in CeRh$_3$B$_2$. The inter-site hybridization of the $\ket{J=\frac{5}{2},J_z=\pm \frac{1}{2}}$ crystal-field state and Ce\,5$d$ band combined with the intra-site Ce\,4$f$--5$d$ exchange creates the strong ferromagnetism, while hybridization between the $\ket{J=\frac{5}{2},J_z=\pm \frac{5}{2}}$ and the B\,$sp$ in the $ab$-plane contributes to the Kondo interaction which causes the moment reduction. This orbital selective coupling explains the unique and seemingly contradictory properties of CeRh$_3$B$_2$.    

\end{abstract}


\maketitle
\section{Introduction}
\subsection{Properties and questions}
Cerium-based intermetallic compounds have been widely studied thanks to the variety of exotic properties that arise from the coexistence of localized Ce\,4$f$ magnetic moments and itinerant electrons. A widely accepted picture for such ``dense'' impurity systems is the Doniach phase diagram \,\cite{Doniach1977} that describes the competition between the Ruderman–Kittel–Kasuya–Yosida (RKKY) coupling and the Kondo-lattice effect as a function of the interaction strength $\cal{J}$ between local rare-earth moments and itinerant conduction electrons. In the (weak interaction) RKKY limit, a spin polarization of the itinerant electrons due to $\cal{J}$ mediates an effective ferro- or antiferromagnetic coupling between the local moments which leads to magnetic order below temperatures that grow like $T_\text{RKKY}$\,$\propto$\,$\cal{J}$\,$^2$. With stronger interactions, however, the Kondo effect starts to dominate. Here, the formation of singlets between localized and itinerant electrons starts at a characteristic Kondo temperature $T_\text{K}$\,$\propto$\,$e^{-1/\cal{J}}$ and effectively leads to a screening of the local moments by the conduction electrons, and consequent suppression of magnetic order. The ordering temperature therefore goes through a maximum as a function of $\cal{J}$ and is suppressed completely at the so-called quantum critical point in the region where $T_\text{RKKY}$\,$\approx$\,$T_\text{K}$. At that point superconductivity is often observed. Beyond, in the regime where $T_\text{RKKY} < T_K$  an intermediate valent Kondo state and composite heavy quasiparticles emerging from local 4$f$ and itinerant conduction electrons determine the near ground state properties\,\cite{Hilbert2007,Wirth2016}.

\begin{figure}[]
    \centering
    \includegraphics[width=0.9\columnwidth]{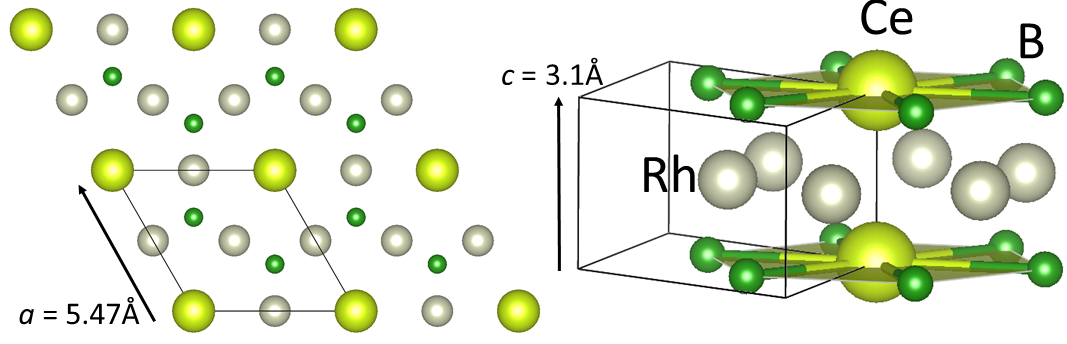}
    \caption{Crystal structure of CeRh$_3$B$_2$, showing the hexagonal $ab$ plane on the left and the alternating layers of Rh and Ce surrounded by B on the right. The room temperature lattice constants are taken from\,\cite{Langen1987}.}
    \label{structure}
\end{figure}

The majority of cerium compounds orders antiferromagnetically, but there are also a fair number of ferromagnetic cerium based compounds. They all have in common that the ordering temperatures tend to be lower than 20\,K\,\cite{Miyahara2018}. CeRh$_3$B$_2$ is the most extreme exception; it orders ferromagnetic  at $T_\text{C}$\,=\,115\,K \cite{Dhar1981,Galatanu2003}, which is by far the highest magnetic ordering temperature recorded for Ce-based intermetallic compounds. Its $T_\text{C}$ is even higher than that of  GdRh$_3$B$_2$, ($T_\text{C}$\,=\,91-105\,K \cite{Malik1983}). It is actually two orders of magnitude larger than expected from de Gennes scaling, which works well for localized moments within the rare earth (RE) series. CeRh$_3$B$_2$ is exceptional in another respect as well. It forms in the hexagonal CeCo$_3$B$_2$-type crystal structure (P6/mmm), that consists of alternating layers of Rh and Ce surrounded by B, and its Ce-Ce distances along the $c$ direction are extraordinary short (see Fig.\,\ref{structure}), shorter than the Hill limit of about 3.5\,{\AA} and shorter than in $\alpha$-Ce ($3.41$\,\AA). The latter material has a non-magnetic, strongly intermediate valent ground state\,\cite{Hill1970,Allen1992}. The $a$ parameter in CeRh$_3$B$_2$ exhibits a shallow minimum as function of temperature at about 200\,K so that its net decrease between 1200 and 5\,K is minor (about 0.2\%). In the same temperature interval, the $c$ parameter decreases by about 2\% and also more drastically than e.g. in the La or Pr compound of the family\,\cite{Langen1987}.  

Despite its large $T_\text{C}$, the magnetic moments in CeRh$_3$B$_2$ are strongly reduced compared to values of a free Ce$^{3+}$ ion, actually more so than what can be expected within a localized crystal-field model. The saturated moments at low temperature measured by magnetic susceptibility \cite{Galatanu2003} are $0.45$\,$\mu_B$ in the basal plane (with a small in-plane anisotropy) and $0.04$\,$\mu_B$ along the hexagonal $c$ axis\,\cite{Dhar1981,Lawson1987}. The magnetic anisotropy can be related to the crystal structure, characterized by a large crystalline anisotropy (see Fig.\,\ref{structure}). 

There is consensus about the filling of the 4$f$ shell being non-integer ($n_f$\,$\approx$\,0.88) in CeRh$_3$B$_2$\,\cite{Maple1985,Sampa1985,Kitaoka1985,Vija1985,Malik1985,Shaheen1986,Fujimori1990,Allen1990,Takeuchi2004,Sundermann2016} and, according to photoemission, the 
Kondo temperature $T_\text{K}$ is of the order of 400\,K. This is large, although not as large as in superconducting CeRu$_3$B$_2$ with its $n_f$\,$\approx$\,0.76 and $T_\text{K}$\,=2000\,K. Actually, in the substitution series Ce(Rh$_{1-x}$Ru$_x$)$_3$B$_2$, the substitution of Rh by Ru quickly destroys the magnetic order\cite{Maple1985,Allen1990}.

Itinerant magnetism associated with the Rh\,4$d$ bands seemed to be a plausible explanation for the unusual magnetism in CeRh$_3$B$_2$ since it is not obvious how the RKKY interaction in the presence of localized moments could produce such a high ordering temperature. However, many facts contradict this; for example, LaRh$_3$B$_2$ is not ferromagnetic\,\cite{Sampa1985} and the Rh sites do not seem to carry a magnetic moment\,\cite{Alonso1998,Grange1999,Sakurai2003,Ito2014}. Many other models have been put forward\,\cite{Shaheen1985JMMM,Shaheen1985,Sampa1985,Malik1985,Maple1985,Allen1990,Okubo2003,Yamauchi2010,Takegahara1985,Kasuya1987,Kasuya1987book,Kono2006}, models that include the strong crystal-electric field along the $c$-axis and the Kondo effect. Until today, however, it is not so clear which states would be involved to establish the magnetism and which states would contribute to the reduction of the Ce moment. The possible importance of the crystal-electric field has been recognized but the crystal-field scheme has yet not been fully determined despite measurements of the magnetic anisotropy\,\cite{Givord2007magn}, neutron scattering work \,\cite{Givord2004,Givord2007INS} and x-ray absorption efforts\,\cite{Fujimori1990,Yamaguchi1992,Yamaguchi1995,Imada2007}.

In this study we aim to determine the electronic structure of CeRh$_3$B$_2$ including the full 4$f$ crystal-ﬁeld scheme in order to obtain answers to the following questions: which states are responsible for the ferromagnetism with the anomalously high ordering temperature, and why at the same time is the saturated magnetic moment so strongly reduced; how strong are the crystal ﬁeld effects, and which states provide the Kondo screening? To this end we performed a combination of spectroscopic experiments and theoretical calculations. We used temperature ($T$) dependent x-ray absorption (XAS) and resonant inelastic x-ray scattering (RIXS), as well as \textit{ab-initio} density functional theory (DFT) calculations. In the past we have shown that XAS with linear polarized light at the Ce $M$-edge gives access to the ground state symmetry when measuring at low $T$, while information about low lying excited states can be obtained by taking XAS data at elevated temperatures\,\cite{Hansmann2008,Willers2009,Rueff2015}. RIXS, on the other hand, has direct access to the crystal-field transition energies and is particularly suited for a compound where large or even $giant$ crystal-field effects are expected. Today, state-of-art beamlines may reach resolutions of 30\,meV at the Ce $M_5$-edge and this resolution does not change over a large energy transfer range. RIXS also has a very good contrast of signal to background, and a strong cross-section for all multiplet states, and, being a two photon process, the selection rules allow excitations beyond those of neutron spectroscopy\,\cite{AmoreseCeRu4Sn6,AmoreseCeRh2Si2,Amorese2022}. The experimental results are then interpreted in light of the relevant projections of the partial density of states and hybridization functions from the DFT calculations. 

\section{Methods}

XAS and RIXS experiments at the Ce $M$-edge (3$d^{10}$4$f^1$\,$\rightarrow$\,3$d^{9}$4$f^2$) were performed at the Dragon beamline at the NSRRC (National Synchrotron Radiation Research Center) in Hsinchu, Taiwan and the soft-RIXS beamlines I21 at DIAMOND Light Source in the UK\,\cite{Zhou2022} and ID32\,\cite{ID32} at the European Synchrotron Radiation Facility (ESRF) in France. All experiments were performed on Czochralski-grown single crystalsAppendix\,\ref{app:sample}. At I21 RIXS data were taken at 15, 135, and 295\,K; at ID32 RIXS measurements were performed with smaller temperature intervals, namely 20, 80, 165, 250, and 318\,K. The data were analyzed with full multiplet calculations using the QUANTY code\,\cite{Haverkort2012}. Greater details about experimental set-ups,  data analysis, and the density functional calculation can be found in the Appendix\,\ref{app:setup}. 

\section{Experimental results}
\subsection{XAS}
\begin{figure}[t]
    \centering
    \includegraphics[width=0.9\columnwidth]{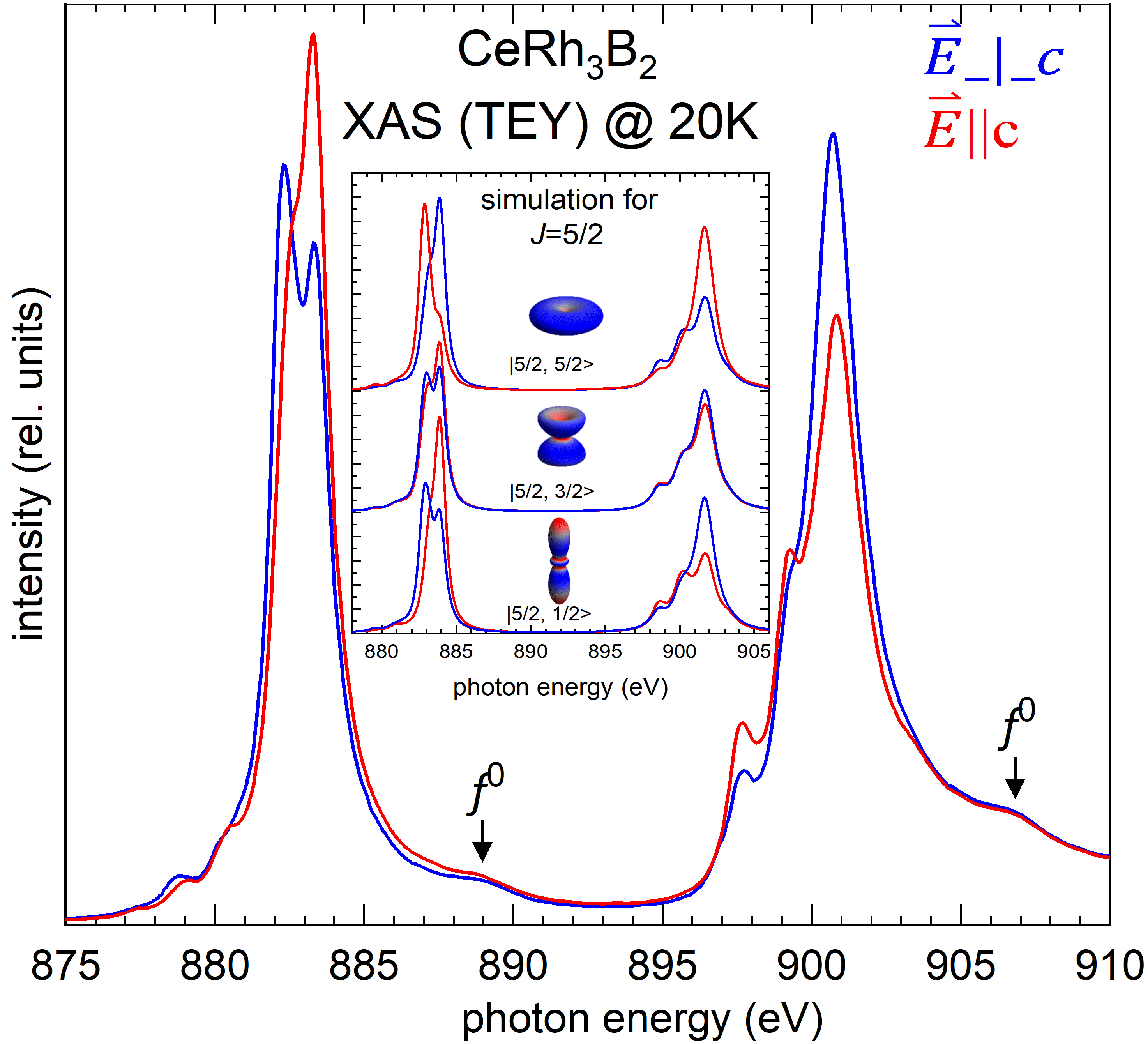}
    \caption{(a) XAS spectra of CeRh$_3$B$_2$ taken at 20\,K with the electric field vector $\vec{E}$ parallel $c$ (red) and in the hexagonal $ab$ plane (blue). The arrows indicate the satellites due to the 4$f^0$ contribution in the ground state. Inset: simulation of XAS spectra of pure $J_z$ states of the $^2$F$_\frac{5}{2}$ multiplet and their respective charge densities.}
\label{XAS}
\end{figure}

\begin{figure}[t]
  \centering
    \includegraphics[width=0.9\columnwidth]{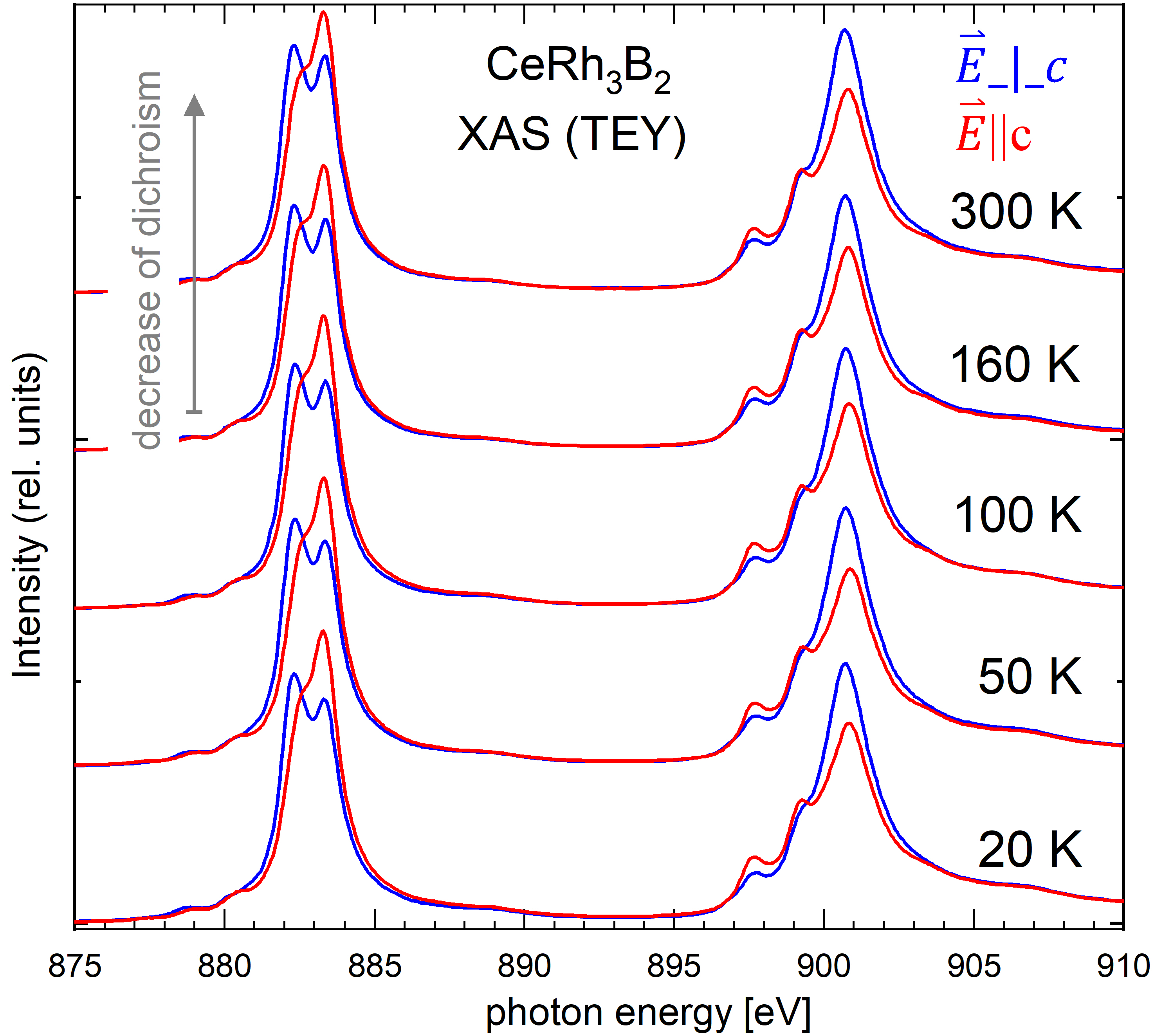}
    \caption{$T$ dependence of the XAS spectra. Electric field vector $\vec{E}$ parallel $c$ (red) and in the hexagonal $ab$ plane (blue). Above 160\,K the dichroism decreases.}
\label{XAS_T}
\end{figure}

\begin{figure*}[]
    \centering
    \includegraphics[width=1.8\columnwidth]{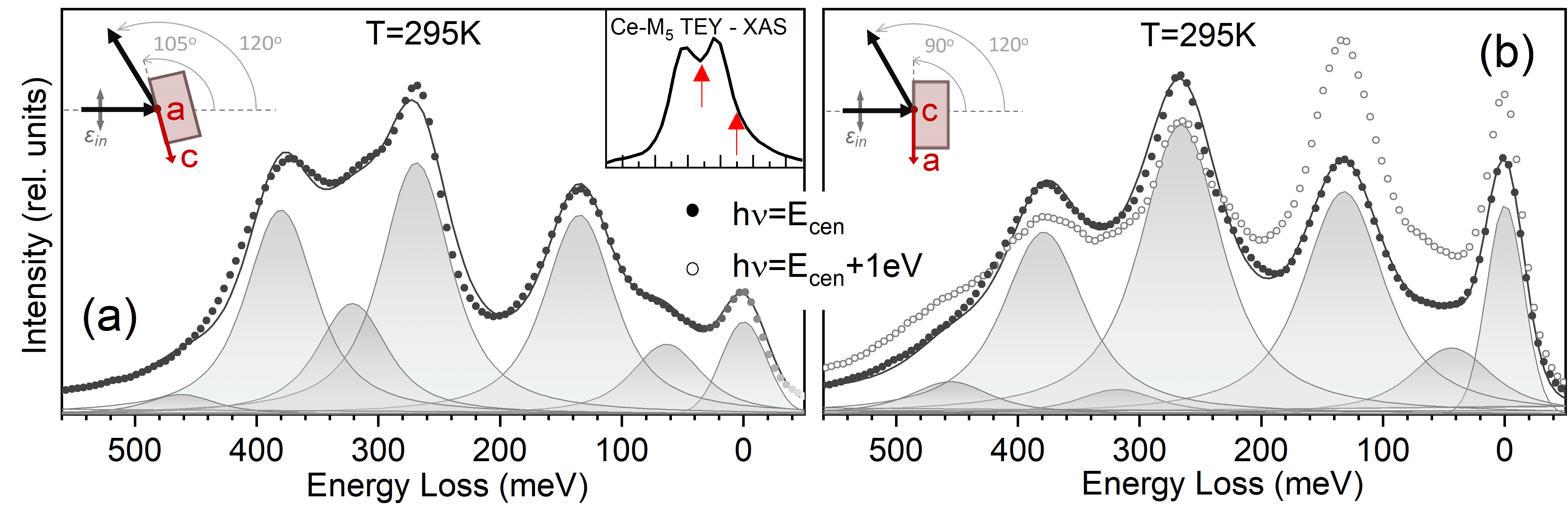}
    \caption{CeRh$_{3}$B$_2$ RIXS data acquired on I21, with the experimental geometries and incident photon energy $h\nu$ indicated in the insets (the isotropic XAS spectrum shown was acquired during the experiment). The lines show the fit of the  $h\nu$\,=\,E$_{cen}$ data, used to determine the crystal-field splitting energies. The inelastic peaks were fitted using Voigt curves, to take into account their intrinsic finite broadening, while the elastic line at 0\,meV, mostly due to non-resonant elastic processes, was fitted with a resolution limited 40\,meV-FWHM Gaussian.}
    \label{RIXS}
\end{figure*}

With XAS at low $T$ we determine the crystal-field wave function of the ground state. Figure\,\ref{XAS} shows the XAS data of CeRh$_3$B$_2$ for the two polarizations $\vec{E}$\,$\parallel$\,$c$ (red) and $\vec{E}$\,$\perp$\,$c$ (blue)  at 20\,K. The crystal-field splitting in CeRh$_3$B$_2$ is expected to be large so that at 20\,K only the ground state is populated. Hence, the XAS spectra at 20\,K are representative for the  crystal-field ground state. The general spectral shape of our XAS data agrees well with the XMCD data of Imada \textit{et al}.\,\cite{Imada2007} who also finds, in addition to the main absorption due the 4$f^1$ configuration in the ground state, a small satellite at the higher energy side of the $M_5$ and $M_4$ absorption lines due to the presence of some 4$f^0$ in the initial state (see arrows in Fig.\,\ref{XAS}). 

We compare the data with a full multiplet calculation (see Appendix\,\ref{app:FM} for details) in order to confirm the anticipated ground state symmetry. This is different from the intentions of Yamaguchi \textit{et al}. who were aiming to find values for the Kondo and Curie temperature when fitting their $N$-edge data within an Anderson impurity model\,\cite{Yamaguchi1995}. 

We recall that the crystal-field Hamiltonian is expressed as a sum of tensor operators $\hat C_k^m$ which transform in the same way as the (renormalized) spherical harmonics $C_k^m(\theta,\phi)=\sqrt{\frac{4\pi}{2k+1}}Y_k^m(\theta,\phi)$. For the hexagonal point symmetry of Ce$^{3+}$ in CeRh$_3$B$_2$ the crystal-field Hamiltonian is 
\begin{align*}
H_{CEF}= &  A_2^0 \hat C_2^0 +  A_4^0 \hat C_4^0 +  A_6^0 \hat C_6^0 +  A_6^6 (\hat C_6^6 +\hat C_6^{-6})\ ,
\end{align*}
where $ A_k^m$ are the crystal-field parameters, expressed using the Wybourne normalization \cite{Wybourne1965}. The $ A_k^m$ have to be determined experimentally. The crystal field splits the $^2$F$_\frac{5}{2}$ and $^2$F$_\frac{7}{2}$ multiplets into three and four Kramers doublets, respectively. Usually, in Ce compounds the crystal-field splitting is much smaller than the spin-orbit splitting, so that for hexagonal point symmetry the eigenstates of the crystal-field Hamiltonian are pure $\ket{J_z}$ states, with the exception of $\ket{J=\frac{7}{2},J_z=\pm \frac{5}{2}}$ and $\ket{J=\frac{7}{2},J_z=\mp \frac{7}{2}}$ that can be mixed by the action of the $ A_6^6 $ parameter. In case of larger crystal-field splittings, however, the two multiplets $^2$F$_\frac{5}{2}$ and $^2$F$_\frac{7}{2}$ begin to intermix, and, former pure $\ket{J_z}$ states, will have the form $\alpha\cdot\ket{J=\frac{5}{2},J_z=\mp \frac{1}{2}}$\,+\,$\sqrt{1-\alpha^2}\cdot\ket{J=\frac{7}{2},J_z=\mp \frac{1}{2}}$. In the $giant$ crystal-field scenario, the entire L--S--J coupling scheme may break down. 

The inset of Fig.\,\ref{XAS} shows the full multiplet simulation for the polarization dependence of the pure  $\ket{\frac{5}{2},J_z}$ states of the $^2$F$_\frac{5}{2}$ multiplet. The polarization dependence of the $\ket{\frac{5}{2},\pm\frac{1}{2}}$ and $\ket{\frac{5}{2},\pm\frac{3}{2}}$ has the same sign as the experiment. The experimental dichroism is much stronger than the simulation for $\ket{\frac{5}{2},\pm\frac{3}{2}}$ so that it can be excluded  as a solution. The data almost resemble the simulation  of the $\ket{\frac{5}{2},\pm\frac{1}{2}}$ state with its cigar shaped charge density, in agreement with the ordered moment in the $ab$ plane and many earlier studies (see e.g. Ref.s\,\cite{Yamaguchi1992,Yamaguchi1995,Givord2004,Givord2007magn,Givord2007INS,Imada2007}). In the experiment, however, some dichroism is missing and this was seen in several repeated measurements performed on freshly cleaved samples. According to Givord \textit{et al}.\,\cite{Givord2007INS} the $^2$F$_\frac{7}{2}$ should contribute to the ground state so that we performed a full multiplet calculation for a mixed ground state, but we find that the multiplet mixing cannot account for the discrepancy between data and simulation (see the simulation of $coherent$ sums\,\cite{Laan1986} of $\ket{\frac{5}{2},\pm\frac{1}{2}}$ and $\ket{\frac{7}{2},\pm\frac{1}{2}}$ in Appendix\,\ref{app:XAS} where we also show the simulations of the pure $\ket{\frac{7}{2},J_z}$ states). We will come back to the missing dichroism in Section IV.

Next we gain some information about excited states from XAS measurements at elevated temperatures (see Fig.\,\ref{XAS_T}). The $T$-dependent data reveal at first no change in dichroism up to 160\,K so that we can exclude low lying states. At 300\,K the dichroism has decreased. We can exclude the partial population of the $\ket{\frac{5}{2},\pm\frac{3}{2}}$ state because it is as high as $\approx$\,150\,meV according to neutron scattering \,\cite{Givord2007INS} and it has the same, although weaker polarization dependence than the ground state (compare inset of Fig.\,\ref{XAS}). The $\ket{\frac{5}{2},\pm\frac{5}{2}}$, on the other hand, has a strong opposite dichroism and, if its energy were of the order of 50\,meV, it would already be partially populated at 300\,K, and thus reduce the net dichroism. It should be noted that due to the selection rules of inelastic neutron scattering, $\Delta J_z=0,\pm 1$, a ground state excitation to a $\ket{\frac{5}{2},\pm\frac{5}{2}}$ state has zero cross-section.  

\subsection{RIXS data at room temperature}

The XAS data have shown that there are no crystal-field states lower than 30\,meV so that at room temperature the population of excited states is still minor. This allows us to start with the discussion of the room temperature data that are well above the Curie temperature, thus avoiding the possible presence of magnon exciations. Figure\,\ref{RIXS} shows two RIXS spectra of CeRh$_3$B$_2$ acquired at beamline I21 at 295\,K.  The respective scattering geometries are sketched in the insets. The full circles refer to incident energies h$\nu$\,=\,E$_{cen}$, the open circles to h$\nu$\,=\,E$_{cen}$\,+\,1\,eV as depicted in the inset with the $M_5$ absorption edge. The data with h$\nu$\,=\,E$_{cen}$\,+\,1\,eV were multiplied by three to compensate for the reduction of the cross section when using an incident energy on the tail of the absorption resonance. The comparison of the three data sets demonstrates the cross-section dependence of the RIXS signal and how it allows the identification of the seven expected peaks: the elastic at zero energy transfer and six excitations from the ground state into the excited crystal-field states, two within the Hund's rule ground state $^2F_{5/2}$ and another four into the excited multiplet $^2F_{7/2}$ .

The energies of the crystal-field states were determined more precisely by fitting the spectra with Fityk\,\cite{Fityk}. The elastic peak at 0\,meV was modeled with a Gaussian function of resolution width. The excitations, visibly broader than the elastic signal, probably due to hybridization effects, were modeled with Voigt profiles. The widths were treated as free parameters with an equally constraint. The width turned out to be 72\,$\pm$\,8\,meV. This way we found for the peak positions 0\, 56$\pm$6, 130$\pm$6, 267$\pm$5, 321$\pm$4, 378$\pm$4 and 464$\pm$10\,meV, where the errors take into account the uncertainties of the peak fitting and the absolute energy scale of the experiment. The estimate of the last peak position suffers of a further uncertainty due to the choice of the line used to model the background. 

\subsection{RIXS T-dependence}
\begin{figure}[]
    \centering
    \includegraphics[width=0.9\columnwidth]{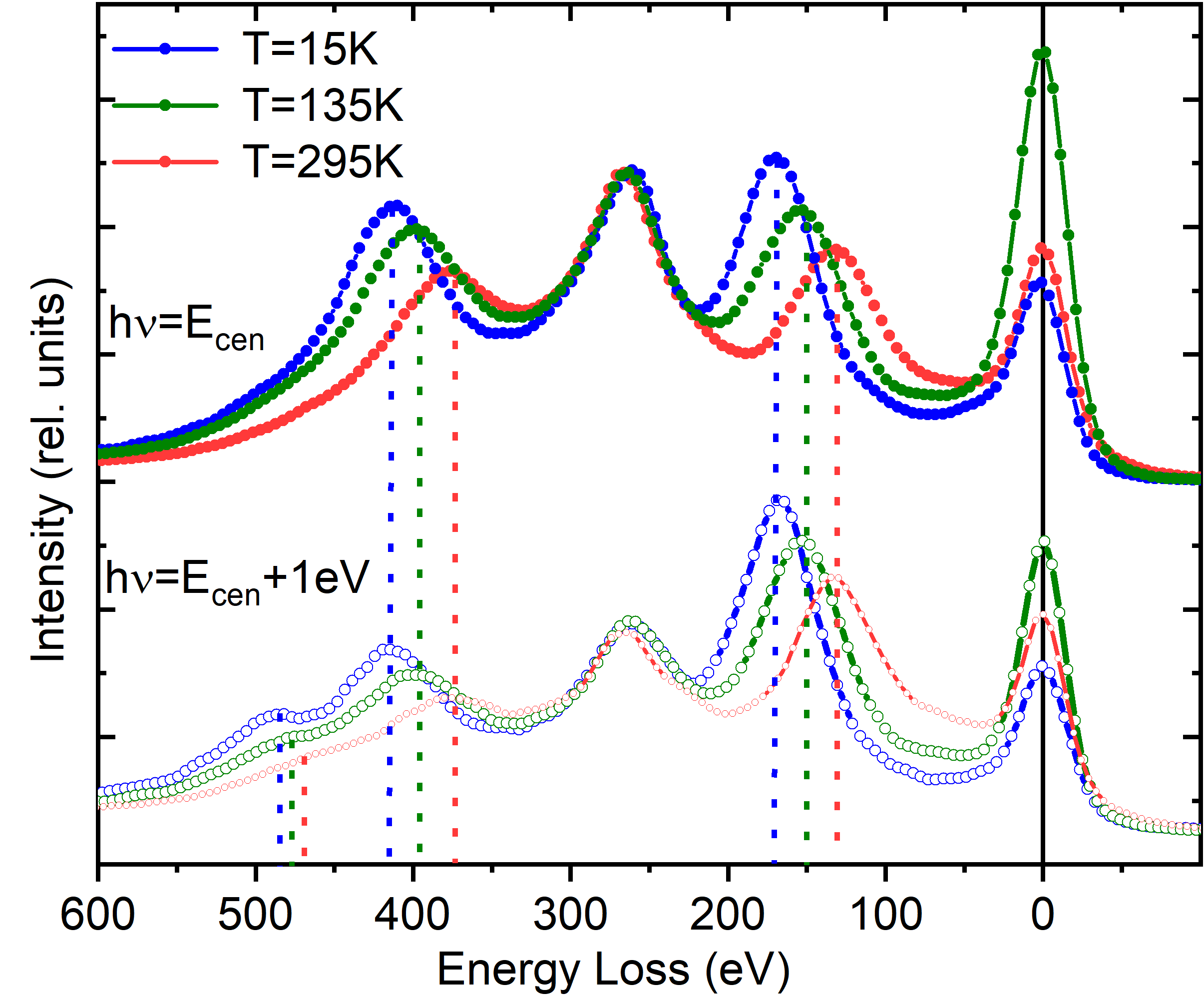}
    \caption{Temperature dependence of CeRh$_{3}$B$_2$ experimental RIXS spectra with incident photon energies $h\nu = \texttt{E}_{cen}$ and geometries as in Fig.\,\ref{RIXS}). }
\label{temp}
\end{figure}

\begin{figure*}[]
    \centering
    \includegraphics[width=1.8\columnwidth]{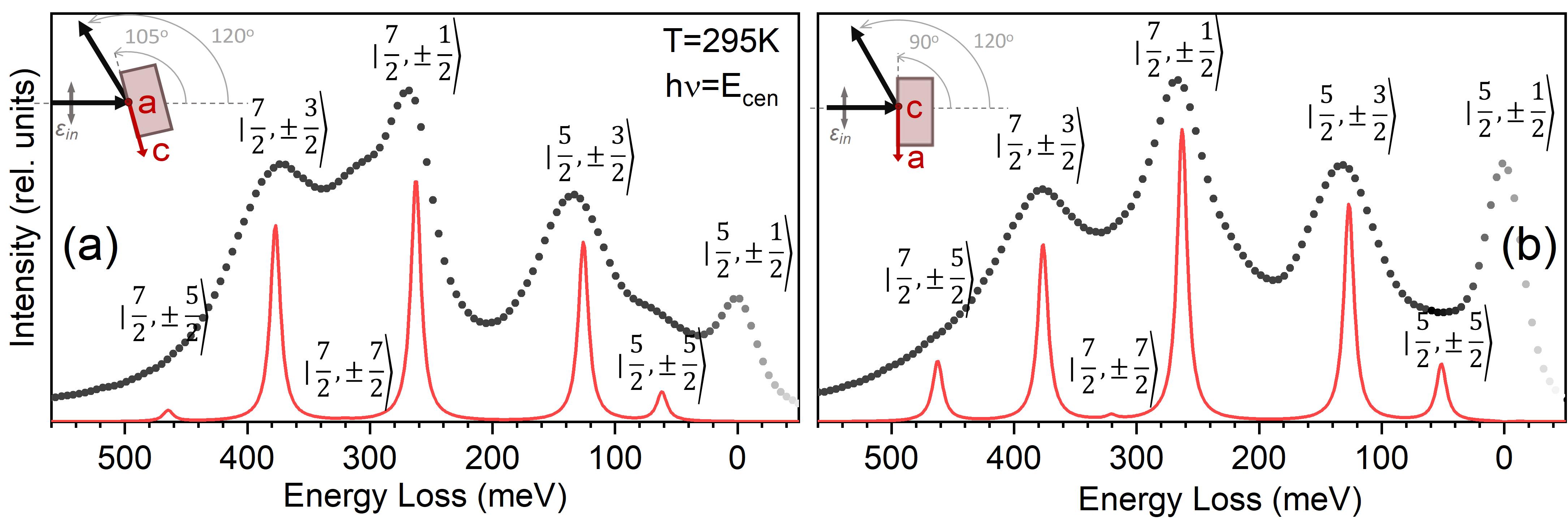}
    \caption{Crystal-field RIXS calculation with reduced broadening (red lines) for the room temperature data (black dots) in Fig.\,\ref{RIXS}. The thermal population of excited states has been considered but only the ground state excitations are indexed because the intensity of non-ground state excitations is negligible. Here the set of parameters is used that describes the data best. It corresponds to the level scheme in Table I, see Appendix\,\ref{app:CEF} for the crystal-field parameters.}
\label{fit}
\end{figure*}

Figure\,\ref{temp} shows the temperature dependence of the RIXS spectra for the same scattering geometry and energies as in the right hand panel of Fig.\,\ref{RIXS}, h$\nu$\,=\,E$_{cen}$ (top) and h$\nu$\,=\,E$_{cen}$\,+\,1\,eV (bottom). Some parts of the spectra exhibit a strong temperature dependence others lesser so. The peak at 260\,meV does not seem to shift with temperature and, most importantly, it does not broaden when entering the ferromagnetic phase so that the Zeeman splitting of the Kramers doublets does not seem to be important. The peaks at about 150\,meV, at about 400\,meV, and also the highest peak are shifted by about 50\,meV to higher energy transfers. In fact, inelastic neutron scattering also observed the strong temperature dependence of the peak at about 150\,meV\,\cite{Givord2007INS}. In RIXS, we further observe that the relative peak intensities are strongly affected by temperature. Data with better resolution and also smaller temperature intervals were taken on ID32 (see Appendix\,\ref{app:ID32}). The smaller temperature intervals show that these spectral changes are gradual with $T$.

\subsection{Crystal-field simulation of RIXS data}
The phase space ($ A_2^0 $,$ A_4^0 $,$ A_6^0 $,$ A_6^6 $) was searched for sets of crystal-field parameters that provide the energy splittings obtained from the empirical fit to the data at room temperature. The $ A_6^6 $ was not scanned because it is less important for the transition energies. Its main impact is the mixing of the $\ket{\frac{7}{2},\pm\frac{5}{2}}$ and $\ket{\frac{7}{2},\mp\frac{7}{2}}$. The search procedure is described in detail in the Appendix\,\ref{app:fit}. It turns out that six clusters of parameters yield the correct energies but different level schemes, two of which have a majority $\ket{\frac{5}{2},\pm \frac{1}{2}}$, two a $\ket{\frac{5}{2},\pm \frac{3}{2}}$, and two a $\ket{\frac{5}{2},\pm \frac{5}{2}}$  ground state. The word $majority$ connotes that there is also some contribution from the $^2F_{\frac{7}{2}}$ multiplet. The RIXS simulations for all six parameter sets using the scattering geometries of the data in Fig.\,\ref{RIXS}, with h$\nu$\,=\,E$_{cen}$, $\phi$\,=\,90$^{\circ}$ and 105$^{\circ}$, are shown in the Appendix\,\ref{app:CEF}. The RIXS calculations have taken into account the appropriate temperature but excitations from excited states turned out to have negligible intensities so that only the ground state excitations are indicated. The parameter sets that correspond to models with a $\ket{\frac{5}{2},\pm \frac{3}{2}}$ or $\ket{\frac{5}{2},\pm \frac{5}{2}}$ ground state can be excluded because they contradict the findings from XAS, and only one of the two models with a $\ket{\frac{5}{2},\pm \frac{1}{2}}$ ground state has a $\ket{\frac{5}{2},\pm \frac{3}{2}}$ excited state at 120\,meV which is close to the energy of the peak seen in inelastic neutron scattering at room temperature\,\cite{Givord2007INS}.  This is the model that will be further discussed. It corresponds to the $red$ simulation in the Appendix\,\ref{app:CEF} and is displayed in Fig.\,\ref{fit}.

The crystal-field model in Fig.\,\ref{fit} with a (majority) $\ket{\frac{5}{2},\pm \frac{1}{2}}$ ground state and the $\ket{\frac{5}{2},\pm \frac{3}{2}}$ state at about 120\,meV, has a majority $\ket{\frac{5}{2},\pm \frac{5}{2}}$ at 54\,meV. Thus, this model agrees with our working hypothesis derived from the temperature dependence of the XAS data. The crystal-field level scheme and corresponding wave functions are listed on in Table\;I (left).  

We then adjusted the crystal-field parameters of the model above to fit the RIXS data from I21 and ID32 at all temperatures (see Appendix\,\ref{app:ID32}). The resulting crystal-field wave functions and energy splittings at 20\,K are also given in Table\,I (right). It is apparent that the multiplet mixing with the $^2F_{\frac{7}{2}}$ multiplet in the ground state has increased considerably.  Table\;III\;and\;IV in the Appendix\,\ref{app:CEF} summarize the corresponding crystal-field parameters and transition energies at all temperatures.

\begin{table*}[]
\centering
\def\arraystretch{1.8}
\caption{Crystal-field wave functions and energy splittings at 320\,K (left) and 20\,K (right) calculated with the crystal-field parameters listed in Table\,III.}
\begin{tabular}{ C{6.5cm} | C{2.1cm} || C{6.5cm} | C{2.1cm} }
T = 320\,K:  $\ket{J,J_{z'}}$                   			& Energy (meV)   & T = 20\,K:  $\ket{J,J_{z'}}$               						& Energy (meV)     \\ 
\hline
\hline
0.998$\ket{\frac{5}{2},\pm\frac{1}{2}} \pm  0.061\ket{\frac{7}{2},\pm\frac{1}{2}}$ 	       	& 0   & 0.974$\ket{\frac{5}{2},\pm\frac{1}{2}} \pm  0.228\ket{\frac{7}{2},\pm\frac{1}{2}}$ &                         0 			\\
\hline
0.969$\ket{\frac{5}{2},\pm\frac{5}{2}} \pm  0.150 \ket{\frac{7}{2},\pm\frac{5}{2}}$	     	& 54  &	0.972$\ket{\frac{5}{2},\pm\frac{5}{2}} \pm  0.167 \ket{\frac{7}{2},\pm\frac{5}{2}}$							& 66			\\ 		
\hline
0.922$\ket{\frac{5}{2},\pm\frac{3}{2}} \mp  0.387 \ket{\frac{7}{2},\pm\frac{3}{2}}$ 		& 120 & 0.928$\ket{\frac{5}{2},\pm\frac{3}{2}} \mp  0.372 \ket{\frac{7}{2},\pm\frac{3}{2}}$ 						& 166			\\ 
\hline
0.992$\ket{\frac{7}{2},\pm\frac{1}{2}} \mp 0.061\ket{\frac{5}{2},\pm\frac{1}{2}}$   		& 262 &	0.974$\ket{\frac{7}{2},\pm\frac{1}{2}} \mp 0.228\ket{\frac{5}{2},\pm\frac{1}{2}}$   							& 270 		\\
\hline
0.977$\ket{\frac{7}{2},\pm\frac{7}{2}}\pm 0.054\ket{\frac{7}{2},\pm\frac{5}{2}} \pm 0.205\ket{\frac{5}{2},\pm\frac{5}{2}}$		& 323 & 0.985$\ket{\frac{7}{2},\pm\frac{7}{2}}\pm 0.024\ket{\frac{7}{2},\pm\frac{5}{2}} \pm 0.171\ket{\frac{5}{2},\pm\frac{5}{2}}$						& 321     \\ 
\hline
0.922$\ket{\frac{7}{2},\pm\frac{3}{2}} \mp 0.387\ket{\frac{5}{2},\pm\frac{3}{2}}$			& 375 & 0.928$\ket{\frac{7}{2},\pm\frac{3}{2}} \mp 0.372\ket{\frac{5}{2},\pm\frac{3}{2}}$								& 413  \\    
\hline
0.987$\ket{\frac{7}{2},\pm\frac{5}{2}} \pm 0.083\ket{\frac{7}{2},\pm\frac{7}{2}} \pm 0.138\ket{\frac{5}{2},\pm\frac{5}{2}}$		& 466   &  0.986$\ket{\frac{7}{2},\pm\frac{5}{2}} \pm 0.052\ket{\frac{7}{2},\pm\frac{7}{2}} \pm 0.161\ket{\frac{5}{2},\pm\frac{5}{2}}$					& 497      \\ 
\end{tabular}                                         
\end{table*}

\section{Discussion}

\subsection{Orbital quenching}
We find that the total crystal-field splitting of the lower mulitplet amounts to about 150\,meV, thus setting the scale for the strength of the crystal-field potential in CeRh$_3$B$_2$. Although larger than the usual values of $\leq$50\,meV, it is still significantly smaller than the spin-orbit interaction separating the two multiplets ($\approx$\,280meV). We thus do not confirm the $giant$ crystal-field scenario by Givord \textit{et al.} with a strongly multiplet mixed ground state (10\%) and strongly reduced in-plane moment as a consequence\,\cite{Givord2007INS,Givord2007magn}.

The crystal-field splitting is nevertheless large enough to cause some mixing of the two multiplets. Comparing the wave functions at room temperature and low $T$ in Table\,I shows that the multiplet mixing increases upon cooling. The 0.4\% contribution of the higher multiplet in the ground state at room temperature has increased to about 5\% at 20\,K. This mixing is not large enough to cause a complete breakdown of the L-S-J coupling scheme, but it modifies the orbital and spin moments of the ground state. 
\begin{table}[]
\def\arraystretch{1.8}
\caption{Moments}
\begin{tabular}{c|c|c|c|c|c}
\hline 
mixing of GS    &  $a^2$                &  $b^2$                 & $\braket{L_z}$ & $\braket{S_z}$ & $\mu_{ab}$ [$\mu_B$]\\
\hline\hline
0               &  $\frac{3}{7}$=0.4286 &  $\frac{4}{7}$=0.5714  & 0.5714         & $\mp$0.072     &    1.29\\  
\hline
0.4\% at 320\,K &  0.489                &  0.511                 & 0.511          &  $\mp$0.011    &    1.24\\
\hline
5\% at 20\,K    &  0.656                &  0.344                 & 0.344          &  $\pm$0.156    &    0.99\\
\hline              							
                                              
\end{tabular}                                         
\label{}
\end{table}

In general, a $\ket{\frac{5}{2},\pm\frac{1}{2}}$ ground state can be written as $a\ket{L_z=0,S_z=\pm \frac{1}{2}}+b\ket{L_z=\pm 1, S_z=\mp\frac{1}{2}}$ with $a^2$\,+\,$b^2$\,=\,1. In Table\;II, the coefficients that we find from the RIXS analysis and the resulting expectation values $ \braket{L_z}$\,=\,$\pm b^2$ and $\braket{S_z}$\,=\,$(\pm a^2$\,$\mp$\,$b^2)$\,$\times$\,$\frac{1}{2}$, as well as the magnetic moments in the basal $ab$ plane are compared to the values in the absence of multiplet mixing. Note, in the absence of multiplet mixing $ \braket{L_z}$ and  $ \braket{S_z}$ have opposite signs. At room temperature the expectation values are similar to the unmixed case, but at 20\,K the multiplet mixing causes an increase of the $\ket{L_z=0,S_z=\pm \frac{1}{2}}$ part in the ground state, with a consequent reduction of $\braket{L_z}$ and with spin and orbital moments having the same sign. We find therefore a partial quenching of the angular moment induced by the increasing crystal-field strength with decreasing temperature, but to a lesser extent than what was discussed in the frame work of  bandstructure calculations \cite{Takegahara1985,Kasuya1987,Kasuya1987book,Eriksson1989,Yamauchi2010}. We also notice that these expectation values partially agree with the XMCD work of Imada \textit{et al.}\,\cite{Imada2007} and the Anderson impurity model analysis of XLD and XMCD data of Yamaguchi \textit{et al}.\,\cite{Yamaguchi1995}, namely that along $c$, the orbital moment $\mu_\text{orb}$\,=\,$L_z$\,$\mu_{B}$ is of the order of, but slightly larger, than the spin moment $\mu_\text{spin}$\,=\,2\,$S_z$\,$\mu_{B}$. The simulation with a magnetic field applied along $a$ gives a ratio  $-\mu_\text{orb}/\mu_\text{spin}$\,$\approx$\,2.5 which is strongly reduced if compared with the standard value of 4 for a $^2F_\frac{5}{2}$ multiplet, and in agreement with magnetization\,\cite{Givord2007magn} and Compton \cite{Yanouac1998,Sakurai2003} measurements.

To summarize, the unbalance of spin and orbital moments can partially account for the moment reduction in the basal plane (see Table\,II) but not for the entire reduction to 0.45$\mu_B$ as suggested in the giant crystal-field scenario proposed by Givord \textit{et al.}\,\cite{Givord2007magn,Givord2007INS}.

\subsection{Kondo effect}
Our \textit{full--multiplet--crystal--field--only} analysis provide a reasonable description of the experimental XAS and RIXS data although there are some discrepancies in the low $T$ dichroism of the XAS spectra and transition matrix elements in the RIXS data, respectively. We believe that these deviations are due to Kondo-type hybridization effects \cite{Kasuya1987book,Jo1990,Eriksson1989,Yamaguchi1995} that are not accounted for in our simulation. A hybridization-induced mixing of crystal-field levels, as discussed for example in literature\,\cite{Wissgott2016,AmoreseCeRu4Sn6,Amorese2020} 
could enhance some of the cross-sections that appear too small in the single-ion approach with its pure $\ket{J_z=\pm\frac{1}{2}}$ ground state. 

We start with the discussion of the missing dichroism in the XAS data at 20\,K that cannot be explained by thermal population of excited states because the first excited state is as high as 66\,meV (see Table\,I). However, such a high lying state can be mixed into the ground state when the Kondo temperature is sufficiently high, and this is indeed the case for CeRh$_3$B$_2$ which has a Kondo temperature of 400\,K\,\cite{Maple1985,Allen1990}. Hence, the $\ket{\frac{5}{2},\pm \frac{5}{2}}$ state at 66\,meV will contribute to the ground state and thus reduce the net polarization because it has a strong and opposite polarization dependence (see inset of Fig.\,\ref{XAS}). 

A more detailed comparison of data and RIXS calculation in Fig.\,\ref{fit} shows that the intensities for transitions with $\Delta J_z$\,=\,$\pm$2 are underestimated in the simulation, and the transition with $\Delta J_z$\,=\,$\pm$3 should have no cross-section at all but it exists in the experiment. These are the transitions into (majority) $\ket{\frac{5}{2},\pm \frac{5}{2}}$, $\ket{\frac{7}{2},\pm \frac{5}{2}}$, and $\ket{\frac{7}{2},\pm \frac{7}{2}}$ states. This cannot be fixed by adjusting the $A_6^6$ parameter that mixes the $\ket{\frac{7}{2},\pm \frac{5}{2}}$ and  $\ket{\frac{7}{2},\mp \frac{7}{2}}$. The thermal occupation of the excited states is also taken care of in the simulation so that something else must be missing in the  calculation. We believe that also here we see the impact of the Kondo effect, namely the Kondo-induced mixing of the 
first excited state into the ground state. The presence of some $\ket{\frac{5}{2},\pm \frac{5}{2}}$ in the ground state would then allow $\Delta J_z$\,=\,0 and $\pm$1 transitions into the before mentioned states, and these transitions usually have a stronger cross-section than $\Delta J_z$\,=\,$\pm$2 transitions\,\cite{Amorese2022}. This explanation is supported by revisiting the inelastic neutron data of Givord \textit{et al}. In their spectra (see Fig.\,2 in\,\cite{Givord2007INS}). some unidentified scattering occurs at about 60\,meV, i.e. at the energy of the $\ket{\frac{5}{2},\pm \frac{5}{2}}$ state, and this transition can only be seen with neutrons when the ground state has components different from $J_z$\,=\,$\pm$1/2.

\subsection{Low impact of Rh\,4$d$ states}
\begin{figure}[]
    \centering
    \includegraphics[width=1.0\columnwidth]{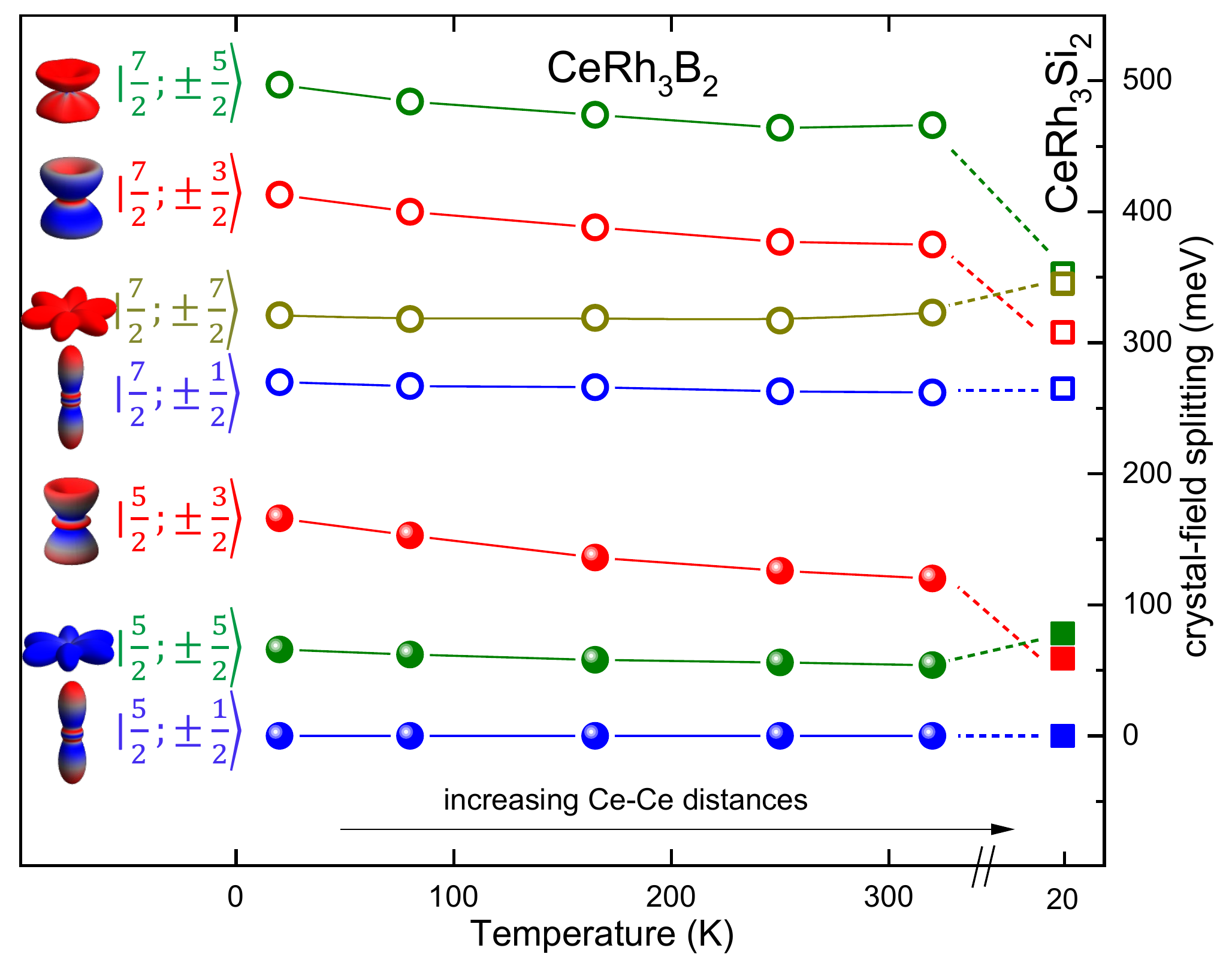}
    \caption{Crystal-field splitting energies of CeRh$_3$B$_2$ as function of temperature (20-320\,K), and for comparison of CeRh$_3$Si$_2$\,\cite{Amorese2022}). The $c$ lattice parameter increases in CeRh$_3$B$_2$ from low $T$ to 300\,K, and even further to pseudo-hexagonal CeRh$_3$Si$_2$ at 20\,K. The symmetries of the states are indexed by their majority $\pm J_z$ and the respective charge densities are shown next to it. Note, the $J_z$\,=\,$\pm$5/2, and $\pm$7/2 states are not rotationally invariant because of the (small) admixtures of $J_z$\,=\,$\pm$5/2 and $\mp$7/2 (see Table I and II). }
\label{energies}
\end{figure}

We next discuss the temperature dependence of the crystal-field energies. They are summarized in Figure\,\ref{energies} (see also Table\,IV). Also here the states are indexed according to their majority $\ket{J ,\pm J_z}$ contribution. The corresponding charge densities are shown on the right. It turns out that the energy difference of ground state and the states with charge densities in the basal plane or along the $c$ axis remain unaffected by cooling and the shortening of the $c$ axis that goes along with it.  This is very different for the energies of the $\ket{\frac{5}{2}, \pm \frac{3}{2}}$, $\ket{\frac{7}{2}, \pm \frac{3}{2}}$, $\ket{\frac{7}{2}, \pm \frac{5}{2}}$ states that have charge densities extending out of the hexagonal $ab$-plane, away from the $c$ direction, and towards the Rh ions (see Fig.\,\ref{structure}). The energy of these states increases upon cooling as much as 50\,meV. This effect  becomes even more apparent when comparing the crystal-field level scheme of CeRh$_3$B$_2$ with that of pseudo-hexagonal CeRh$_3$Si$_2$\,\cite{Amorese2022}. The latter compound is a low $T_\text{K}$ Kondo system that orders antiferromagneticly with $T_N$\,$\leq$\,4.7\,K\,\cite{Pikul2010}. In CeRh$_3$Si$_2$ the Ce--Ce distances are distinctively larger than in CeRh$_3$B$_2$ (see lattice constants in Appendix\,\ref{app:lattice}) and the Ce orbitals that point towards the Rh ions are lower in energy (see Fig.\,\ref{energies}). Hence, the shorter the Ce--Ce distances (when going from CeRh$_3$Si$_2$ to CeRh$_3$B$_2$ at room temperature, and upon cooling in CeRh$_3$B$_2$ to enter the ferromagnetic regime), the higher in energy the Ce orbitals pointing to the Rh ions become i.e. these orbitals become less important for the ground state properties. In short, Rh states play a marginal role to induce the ferromagnetism and/or Kondo hybridization. 

It should be pointed out that the Zeeman splitting cannot be responsible for the temperature dependence of the crystal-field states. The full multiplet calculation with an exchange-field in the basal plane shows that only the two $\ket{J, \pm \frac{1}{2}}$ states are Zeeman split but their energies are not subject to changes with temperature nor does the line width or shape of the $\ket{\frac{7}{2}, \pm \frac{1}{2}}$ change with $T$(see Fig.\,\ref{temp}).

\subsection{Impact of 4$f$\,$\ket{J_z = \pm \frac{1}{2}}$ states}
\begin{figure}[H]
    \centering
   \includegraphics[width=0.99\columnwidth]{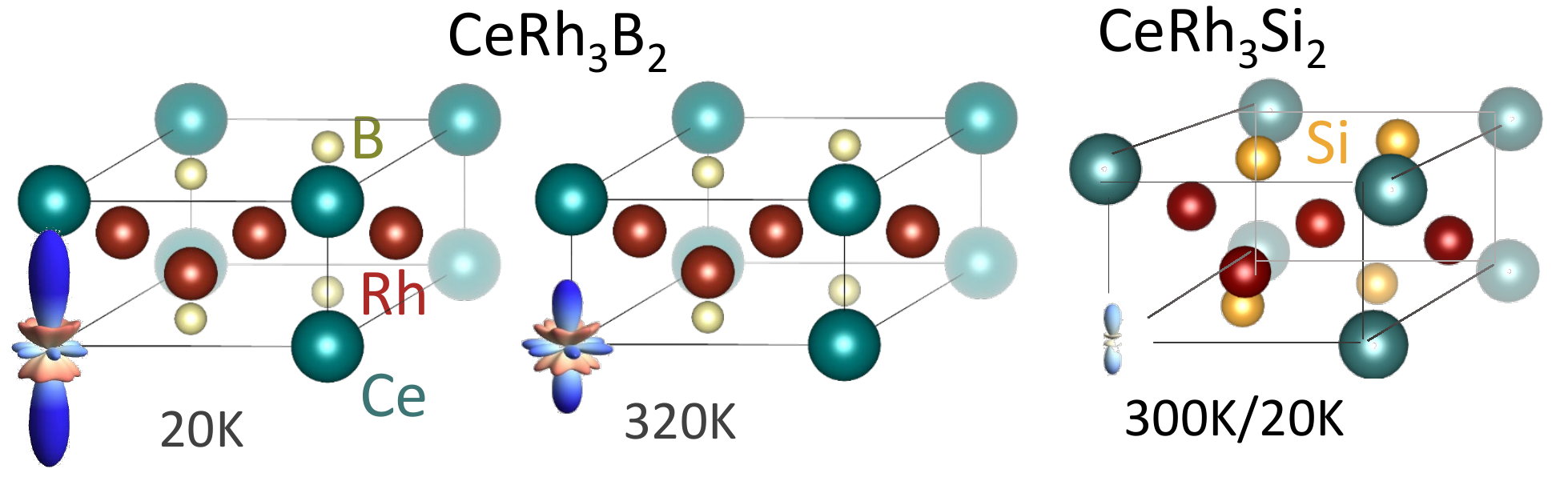}
    \caption{Effective angular part of the crystal-field potential as calculated from weighing the sums of the spherical harmonics with the respective crystal-field parameters. The potentials, depicted on the Ce sites in the hexagonal unit cell of CeRh$_3$B$_2$ at 320 and 20\,K, and in the pseudohexagonal unit cell of CeRh$_3$Si$_2$, have the same relative arbitrary scale. The blue (red) color indicates the directions more (less) favorable for the electrons. The Si compound has the distinctively larger $c$ parameter (see Fig.\,\ref{lattice} in Appendix\,\ref{app:lattice}). }
\label{potential}
\end{figure}
Figure\,\ref{potential} summarizes the effect of the crystal-field potentials over a Ce site in the crystal structure. Shown are the effective angular parts of the crystal-field potentials (in which the expectation values of the radial part are included in the coefficients) using the sum of spherical harmonics $C_k^m(\theta,\phi)$, weighted with the $ A_k^m $ parameters from Table\,III for CeRh$_3$B$_2$ and from Ref.\,\cite{Amorese2022} for CeRh$_3$Si$_2$.  The same (arbitrary) scale within the unit cell is used for the three plots. The relative sizes cannot be understood intuitively due to the complex and non-linear relationship of the crystal-field parameters and splittings (see Table\,III and IV). The directions more energetically favorable for the 4$f$ electrons are indicated in blue, and the least favorable in red. These plots show that the crystal-field potential stabilizes the $\ket{\pm \frac{1}{2}}$ in CeRh$_3$B$_2$ at low $T$ with respect to room temperature, and especially with respect to the RKKY compound CeRh$_3$Si$_2$. This confirms that the dominant interactions are along the Ce-Ce chains in the $c$ direction.

\subsection{\textit{Ab-initio} DFT calculations}
To substantiate our empirical findings, we performed \textit{ab-initio} DFT calculations. Our aim is to find the material-specific parameters that can be used in the periodic Anderson impurity model (pAIM) for CeRh$_3$B$_2$. The DFT calculations are non-spin polarized and full-relativistic with the Ce $4f$ states included as itinerant bands. We used the full potential local-orbital (FPLO) code \cite{Koepernik1999} with a 12\,$\times$\,12\,$\times$\,12 \textbf{k} mesh for the Brillouin zone (BZ) integration. After full self-consistency, we followed a standard downfolding protocol (see e.g.\,\cite{Haverkort2012}) to project the converged Kohn-Sham states to a Wannier basis which consists of the Ce\,4$f$ $\ket{J,J_z}$ states as well as all Rh\,$4d$, B\,$sp$, and Ce\,$5d$ bands in an energy window from $-10$eV to $10$\,eV around the Fermi level. We thus obtained $\varepsilon_{J_z}$ as the on-site energy of the local Ce\,$4f$ states (including the local crystal-electric field potentials), $\varepsilon_{c,\mathbf{k}}$ which denotes the dispersion relation of the uncorrelated conduction electrons (with the index $c$ labeling the Rh\,$4d$, B\,$sp$, and Ce\,$5d$ orbital character), and $V_{J_z,c}(\mathbf{k})$ which describes the hopping integral between the Ce\,$4f$ and the conduction electrons. These quantities from the downfolding process are input for the periodic Anderson impurity model, see Appendix~\ref{app:pAM}.

From $\varepsilon_{c,\mathbf{k}}$ we get directly the Green's function $G^0_c(\omega,\mathbf{k})$ and the partial density of states $\rho^0_c(\omega,\mathbf{k}) = -1/\pi\text{Im}[G^0_{c}(\omega,\mathbf{k})]$ of the uncorrelated conduction electrons. Here the superscript ''0" denotes that these quantities refer to the uncorrelated part of the Hamiltonian. We note also that the partial density of states of these downfolded bands is not to be confused with the original DFT partial density of states since in DFT the hybridization between the Rh\,4$d$, B\,$sp$, and Ce\,5$d$ bands with the Ce\,4$f$ states is included. We will also make use of the quantity $\Delta_{J_z}(\omega,\mathbf{k}) = \sum_c |V_{J_z,c}(\mathbf{k})|^2 \cdot G^0_c(\omega,\mathbf{k})$, which is the so-called hybridization function and describes the effective hybridization strength of the downfolded Ce\,$4f$ and uncorrelated electron bands as function of momentum and energy. A more concise mathematical description of this procedure is presented in Appendix~\ref{app:pAM}.

\begin{figure}[H]
      \centering
    \includegraphics[width=0.99\columnwidth]{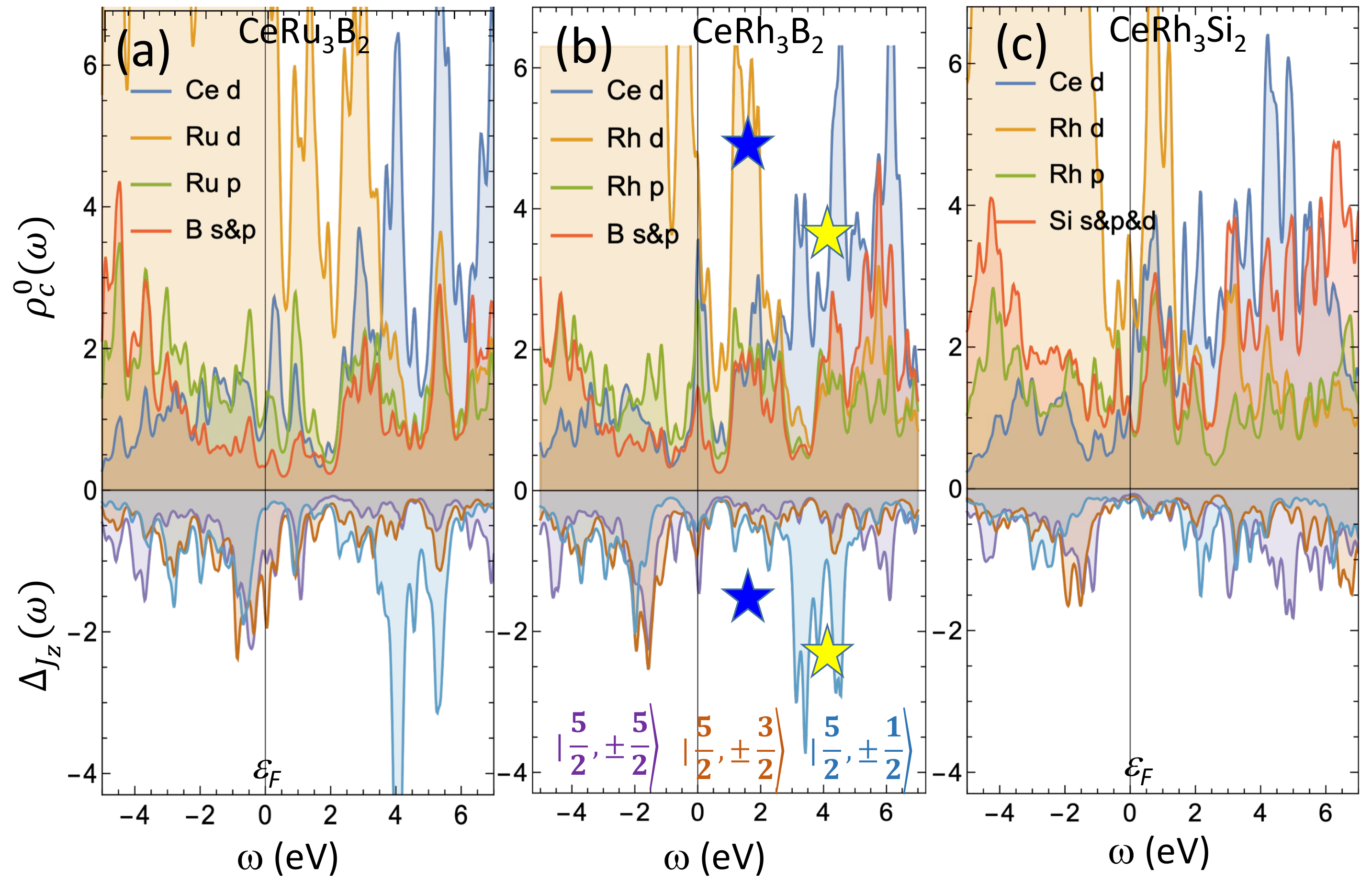}
    \includegraphics[width=0.99\columnwidth]{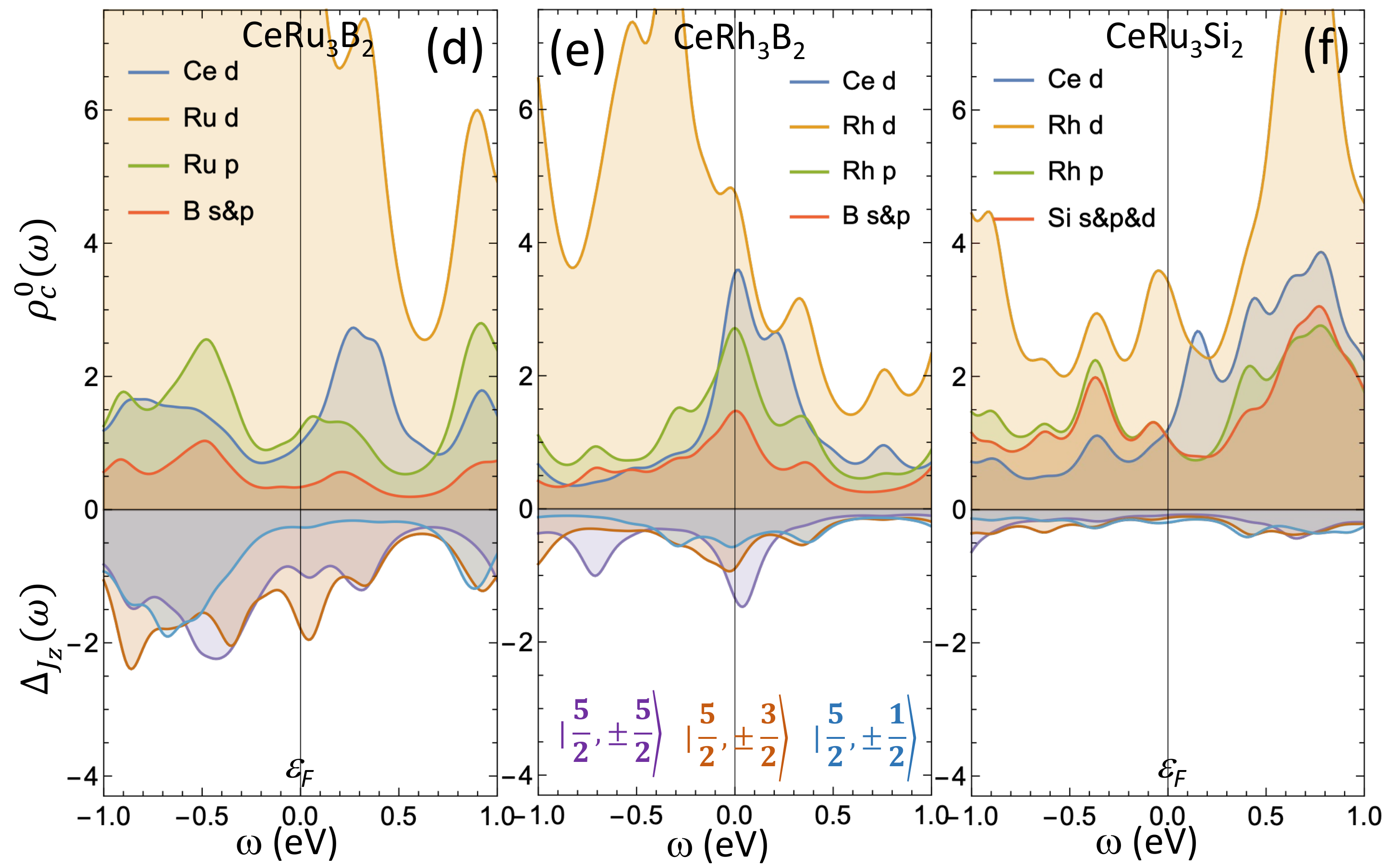}
    \caption{Top of each panel: calculated $\rho^0_c(\omega)$ of the downfolded Ce\,5$d$ states in blue, Rh or Ru\,4$d$ states in orange, Rh or Ru\,4$p$ states in green, and B or Si\,$s$, $p$, and $d$ states in brown. Bottom of each panel: calculated hybridization function $\Delta_{J_z}(\omega)$ of the three lowest Kramer's doublets for CeRu$_3$B$_2$ (left), CeRh$_3$B$_2$ (middle), CeRh$_3$Si$_2$ (right). Panels (a)--(c) display the larger energy scale that is most relevant for the magnetic exchange, while panels (d)--(f) expand the region close to the Fermi level $\epsilon_F$ that is most relevant for the Kondo physics.}
\label{hyb}
\end{figure}

For our analysis we have performed calculations for the strongly intermediate valent superconductor CeRu$_3$B$_2$ ($T_\text{K}$\,$\approx$\,2000K), the high $T_\text{C}$ ferromagnet CeRh$_3$B$_2$ ($T_\text{K}$\,$\approx$\,400\,K) that is subject of the present manuscript, and the pseudo-hexagonal low $T_\text{K}$ RKKY system ($T_N$\,=\,4\,K) CeRh$_3$Si$_2$ (see Fig\,\ref{hyb}\,(a)\,--\,(f)). The two boron compounds have small Ce-Ce distances, and the Si compound has the enlarged $c$ parameter. 

At first we discuss our results in a large energy window of about 10\,eV around the Fermi energy $\epsilon_F$. While the hybridization at these energies is not directly related to the Kondo effect, it is relevant for the magnetic interactions (Fig.\,\ref{hyb}\,(a)\,--\,(c)). The top of the respective panels shows the \textbf{k}-integrated $\rho^0_c(\omega)$ of CeRu$_3$B$_2$ (a), CeRh$_3$B$_2$ (b), and CeRh$_3$Si$_2$ (c) with the downfolded Ce5$d$ states in blue, Rh or Ru\,4$d$ states in orange, Rh or Ru\,4$p$ states in green, and B or Si\,$s$, $p$, and $d$ states in brown. The bottom of the panels shows the \textbf{k}-integrated hybridization functions $\Delta_{J_z}(\omega)$ for the respective $\ket{\frac{5}{2}, \pm \frac{1}{2}}$ (purple), $\ket{\frac{5}{2}, \pm \frac{3}{2}}$ (brown), and $\ket{\frac{5}{2}, \pm \frac{5}{2}}$ (blue) Kramer's doublets.

We find for the two boron compounds, (a) and (b), that the $\ket{\frac{5}{2}, \pm \frac{1}{2}}$ state shows very strong hybridization at energies where the Ce\,5$d$ states also show a very high $\rho^0_c(\omega)$ (see yellow stars). We recall that this is the state that contributes most to the near ground state of CeRh$_3$B$_2$ according to our XAS and RIXS measurements. On the other hand, we cannot identify such overlap of strong hybridization and strong $\rho^0_c(\omega)$ for Rh\,4$d$ (or Ru\,$4d$) states above $\epsilon_F$ (see blue stars). This hints towards an important entanglement of the electrons in the Ce\,5$d$ and Ce\,4$f$\,$\ket{\frac{5}{2}, \pm \frac{1}{2}}$ states for the magnetic interaction that in case of CeRh$_3$B$_2$ leads to a ferromagnetic ground state. This is contrasted by the transition metal 4$d$ electrons that seemingly do not contribute to the formation of magnetic order.  

The above is in agreement with the suggestion of Takegahara \textit{et al.}\,\cite{Takegahara1985}, Kasuya \textit{et al.}\,\cite{Kasuya1987book,Kasuya1987}, and Yamaguchi et al.\,\cite{Yamaguchi1992,Yamaguchi1995} that inter-site Ce--Ce hopping of Ce\,4$f$\,$\ket{\pm \frac{1}{2}}$ ($f_0$ in their nomenclature) and Ce\,5$d$ electrons in combination with intra-atomic 5$d$--Ce\,4$f$\,$\ket{\pm \frac{1}{2}}$ exchange is responsible for the ferromagnetism with high Curie temperature. We do not agree, though, with their suggestion that Ce\,4$f$-Rh\,4$d$ hybridization is important.  

The zoom to the energy window closer to $\epsilon_F$, which is most relevant for the Kondo interaction, highlights the differences of the two boron compounds. In CeRu$_3$B$_2$  (panel(d)), the $\ket{\frac{5}{2}, \pm \frac{3}{2}}$ state, the state with charge densities pointing towards the transition metal ions (here Ru), has a pronounced maximum at $\epsilon_F$ where the Ru\,$4d$ states have a very high $\rho^0_c(\omega)$. Hence, in CeRu$_3$B$_2$ the Kondo effect is triggered by the Ru\,$4d$ and Ce\,4$f$\,$\ket{\frac{5}{2}, \pm \frac{3}{2}}$ electrons and is strong enough to overcome any instability to magnetic order. This is contrasted by CeRh$_3$B$_2$ (panel (e)) where we find  that close to $\epsilon_F$ the hybridization functions are generally reduced, and also the $\rho^0_c(\omega)$ of the   transition metal Rh 4$d$ is much lower. Hence, in CeRh$_3$B$_2$ the magnetic order prevails. This is very much in line with the interpretation of the substitution series Ce(Ru$_{1-x}$Rh$_x$)$_3$B$_2$ by Allen \textit{et al.}\,\cite{Allen1990}.

Nevertheless, also in CeRh$_3$B$_2$ some the Kondo interaction is present and sufficiently strong to reduce the magnetic moments. In contrast to CeRu$_3$B$_2$, we find in CeRh$_3$B$_2$ that the 4$f$\,$\ket{\frac{5}{2}, \pm \frac{5}{2}}$ states with charge densities in the $ab$ plane are most strongly hybridized at $\epsilon_F$ accompanied by peaks in the $\rho^0_c(\omega)$ of B\,$sp$ that are located in the same plane. Hence, in CeRh$_3$B$_2$ it seems that the Kondo effect is mainly triggered by hybridization in the $ab$ plane of the disk-like $\ket{\frac{5}{2}, \pm \frac{5}{2}}$ states with the B\,$sp$ electrons. The hybridization of Ce\,4$f$\,$\ket{\frac{5}{2}, \pm \frac{3}{2}}$ and $\ket{\frac{5}{2}, \pm \frac{1}{2}}$ and Rh\,4$d$ states, in contrast, seems to be significantly less important. This interpretation of the hybridization function is supported by the experimental findings of the presence of the Ce\,4$f$\,$\ket{\frac{5}{2}, \pm \frac{5}{2}}$ in the ground state and the energetically expensive Ce\,4$f$\,$\ket{\frac{5}{2}, \pm \frac{3}{2}}$ states in CeRh$_3$B$_2$. It would be an interesting task for future work to find out where in energy the Ce\,4$f$\,$\ket{\frac{5}{2}, \pm \frac{3}{2}}$ states of CeRu$_3$B$_2$ are in a total energy diagram. 

Finally, CeRh$_3$Si$_2$ (see Fig.\,\ref{hyb}\,(c)\,\&\,(f)) shows an overall reduced hybridization function, completely in line with its antiferromagnetic RKKY (i.e. weak hybridization limit in Doniach's picture) ground state.

\section{Conclusion}
The full crystal-field scheme of CeRh$_3$B$_2$ has been determined with XAS and RIXS. We confirm the Ce\,4$f$\,$\ket{\frac{5}{2}, \pm \frac{1}{2}}$ ground state and find a large but not  $giant$ crystal-field splitting. Consequently, the $J$-$L$-$S$ coupling scheme remains basically intact despite some multiplet mixing and consequent quenching of the orbital moment. The $^2F_{5/2}$--$^2F_{7/2}$ multiplet mixing is not sufficient to explain the reduced magnetic moments.  The interpretation of the temperature dependence of the crystal-field level scheme and the comparison with the level scheme of the RKKY-system CeRh$_3$Si$_2$ shows that the dominant interaction is along the Ce-Ce chains along $c$, and that a contribution of the Rh\,4$d$ electrons to the ground state is unimportant. We further find some Kondo induced mixing of the  Ce\,4$f$\,$\ket{\frac{5}{2}, \pm \frac{5}{2}}$ into the ground state. The experimental findings are used to focus on the relevant projections in the DFT calculations of the uncorrelated conduction bands and the computation of the hybridization functions of the lowest three Kramer's doublets. The calculations also show that the Rh\,4$d$ electrons are not responsible for the ferromagnetic state and that they are of lesser importance for the Kondo interaction. We conclude that the inter-atomic hybridization of the Ce\,4$f$\,$\ket{\frac{5}{2}, \pm \frac{1}{2}}$ and Ce\,5$d$ states in combination with intra-site 4$f$--5$d$ exchange set up the ferromagnetic state, while hybridization of the Ce\,4$f$\,$\ket{\frac{5}{2}, \pm \frac{5}{2}}$ and the B\,$sp$ in the $ab$-plane trigger the Kondo interaction that strongly reduces the magnetic moment. The co-existence of high Curie temperature and high Kondo temperature can thus be traced back to the presence of two different Ce\,4$f$ orbitals, i.e. the $\ket{\frac{5}{2}, \pm \frac{1}{2}}$ and $\ket{\frac{5}{2}, \pm \frac{5}{2}}$, each having their own unique coupling to the different bands in CeRh$_3$B$_2$. 

\section{Appendix}

\subsection{Crystal growth}
\label{app:sample}
The single crystals of CeRh$_3$B$_2$ were grown by Czochralski in a tri-arcs furnace under high quality argon atmosphere and annealed under ultra high vacuum at 950°C for 10 days. The crystals come from the same lab as the samples used in Ref.\,\cite{Givord2004,Givord2007magn,Raymond2010}. Single-crystallinity and orientation were verified with a Laue camera prior to all experiments, XAS and RIXS. 

\subsection{Experimental set-up}
\label{app:setup}
\subsubsection{XAS}
The $M$-edge (3$d^{10}$4$f^1$\,$\rightarrow$\,3$d^{9}$4$f^2$) XAS experiments on CeRh$_3$B$_2$ were performed at the Dragon (bending magnet) beamline at the NSRRC (National Synchrotron Radiation Research Center) in Hsinchu, Taiwan. Measurements were performed with the electric-field vector being $\vec{E}$\,$\parallel$\,$c$ and $\vec{E}$\,$\perp$\,$c$. This was achieved by using a rotatable sample holder. For all measurements the sample was rotated four times by 90$^{\circ}$, so that for each polarization two equivalent positions were measured. The comparison of these spectra ensured a proper orientation of the sample with respect to the polarization vector. The spectra were recorded with the total electron yield (TEY) method and normalized to the incoming flux that was measured on a gold mesh. The energy resolution was set to 0.4\,eV and the degree of linear polarization was $\approx$98\%\,\cite{Hansmann2008}.  

\subsubsection{RIXS}
The $M$-edge RIXS process encounters the absorption process 3$d^{10}$4$f^1$\,$\rightarrow$\,3$d^{9}$4$f^2$, immediately followed by the corresponding 4$f$\,$\rightarrow$\,3$d$ resonant emission. After this two photon process, an excitation can be left in the sample in the final state. For example, the electron in the 4$f$ shell of the Ce$^{3+}$ ion can occupy an excited crystal-field state, and the corresponding excitation energy will be detectable as an energy loss of the scattered photon. Each peak in the energy loss spectrum will therefore correspond to a crystal-field excitation with that energy. The RIXS cross section dependence on the incident photon energy (across the $\approx$ 2\,eV-wide range of the $M_5$ edge) and on the scattering geometry provides the possibility to enhance or suppress excitations. Horizontal incident polarization ($\pi$) was used to minimize the contribution of elastic scattering around 0\,eV. The incident energies h$\nu$ were chosen at the centre of the $M_5$-edge (h$\nu$\,=\,E$_{cen}$) or 1 eV above (E$_{cen}$\,+\,1\,eV), the analyzer was positioned at 120$^{\circ}$ away from forward scattering, and the sample was turned either perpendicular (90$^{\circ}$) to the incoming beam or to 105$^{\circ}$ (see insets in Fig.\,\ref{RIXS}). 

RIXS experiments were preformed at the two state-of-the-art high resolution soft-RIXS beamlines I21 beamline at Diamond Light Source in the UK\,\cite{Zhou2022} and ID32 at European Synchrotron Radiation Facility (ESRF) in France\,\cite{ID32}. At I21 data were taken at 15, 135, and 295\,K, at ID32 measurements were performed with smaller temperature intervals, namely 20, 80, 165, 250, and 318\,K. At both beamlines, the CeRh$_3$Si$_2$ single crystals were post-cleaved in vacuum at room temperature just prior to inserting the sample in the measurement chamber (P\,$\approx\,10^{-10}$\,mbar). The resolution and zero energy were determined by measuring a carbon tape before and after the acquisition of the spectra. At ID21 the resolution was $\approx$40\,meV and at ID32 is was optimized to $\approx$30\,meV. The I21 and I32 data are presented in different ways: the data from I21 are shown with the intrinsic smoothing coming from the use of CCDs sensors for x-ray detection\,\cite{CCD} whereas the ID32 data are presented in the so-called single photon counting algorithm where this intrinsic smoothing has been removed. The latter is necessary to achieve the highest experimental resolution of 30\,meV and to extract realistic statistical error bars\,\cite{GhiringhelliSAXES}.

\subsection{Full multiplet calculation}
\label{app:FM}
The XAS and RIXS data were simulated with full multiplet calculations using the QUANTY code\,\cite{Haverkort2012}.  In the calculations, the Slater integrals for the electron--electron interactions in the final (XAS) or intermediate (RIXS), respectively, $3d^94f^2$ state, as well as the $3d$ and $4f$ spin-orbit parameters, were obtained from the Robert D. Cowan's Atomic Structure Code\,\cite{CowanBook}. Reduction factors take care of configuration interaction effects that are not included in the Hartree-Fock scheme. For the RIXS calculation the parameters were optimized to best reproduce the peak positions. The $4f$ spin-orbit parameter was reduced to about 86\% of the value calculated for an isolated ion. The $F_{ff}$ integrals for the intermediate RIXS state were scaled to 55\% and the $F_{df}$ and $G_{df}$ integrals to 75\% of their calculated value, having been tuned by fitting the isotropic XAS spectrum. The same values were used for calculating the XAS spectra of the pure $\ket{J,J_z}$ states. 

\subsection{XAS simulation of pure $J_z$ states}
\label{app:XAS}
\begin{figure}[]
      \centering
    \includegraphics[width=0.90\columnwidth]{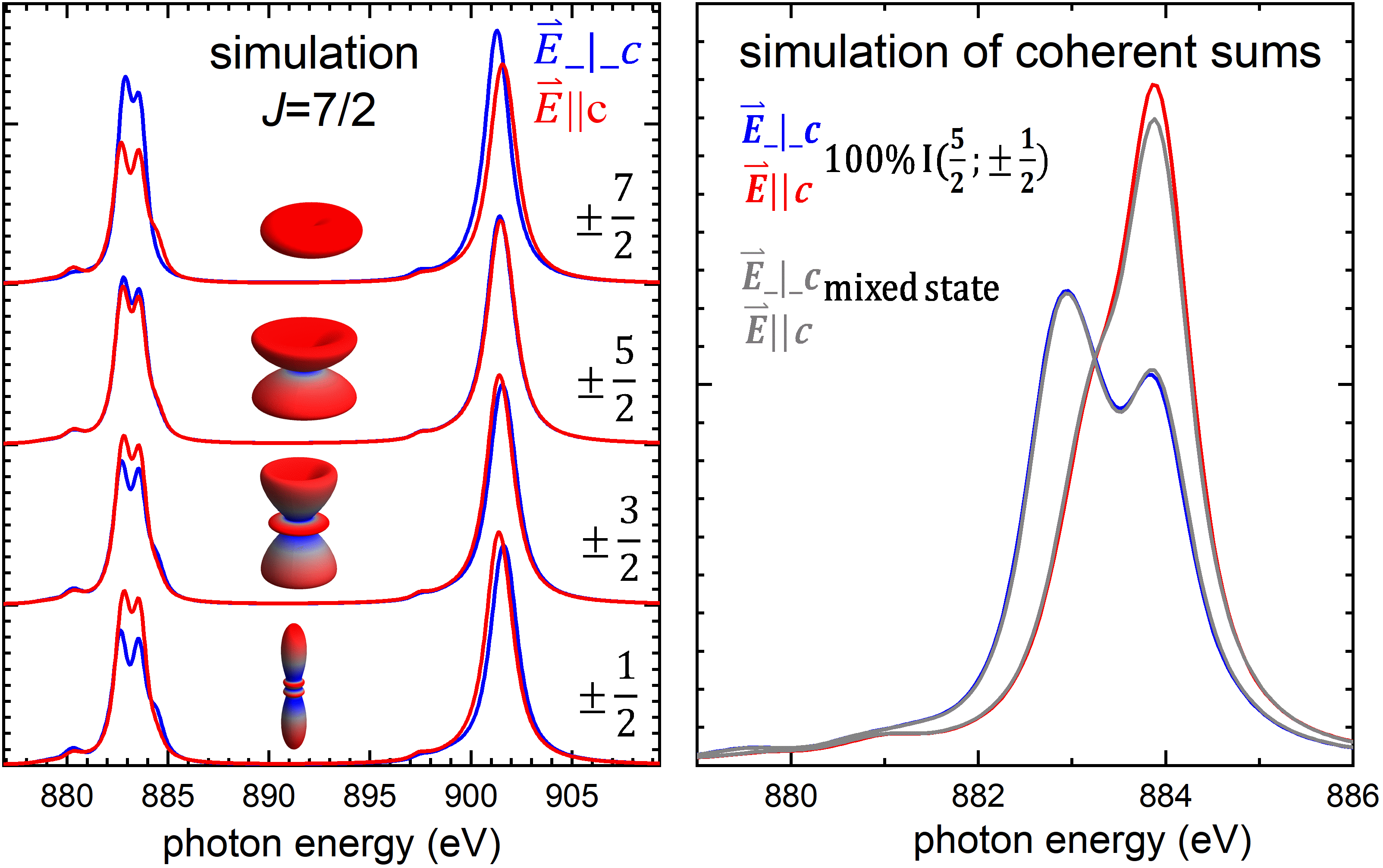}
    \caption{Simulation of pure $\ket{\pm J_z}$ states of the $^2$F$_\frac{7}{2}$ (right) and comparison of the simulated spectra of the pure $\ket{\frac{5}{2},\pm \frac{1}{2}}$ state and the spectra of the multiplet mixed ground state at 20\,K - based on the full multiplet calculation with crystal-field parameters in Table\,III at 20\,K (left).}
\label{sim}
\end{figure}

The left of Fig.\,\ref{sim} shows the simulations for the XAS spectra of the pure $\ket{\frac{7}{2},\pm J_z}$ states. On the right, we compare the simulation for a pure $\ket{\frac{5}{2}, \pm\frac{1}{2}}$ and for the multiplet mixed ground state at 20\,K, $\alpha\cdot\ket{\frac{5}{2},\mp \frac{1}{2}}$\,+\,$\sqrt{1-\alpha^2}\cdot\ket{\frac{7}{2},\mp \frac{1}{2}}$ with 1\,$-$\,$\alpha^2$\,=\,5\% that we obtain from the crystal-field analysis of the RIXS spectra at 20\,K (see Table\,I). Although the multiplet mixing does reduce the dichroism, this reduction is not sufficient to explain the XAS spectra at low $T$ (see Fig.\,\ref{XAS}). 

\subsection{RIXS data from ID32} 
\label{app:ID32}
\begin{figure}[]
      \centering
    \includegraphics[width=0.8\columnwidth]{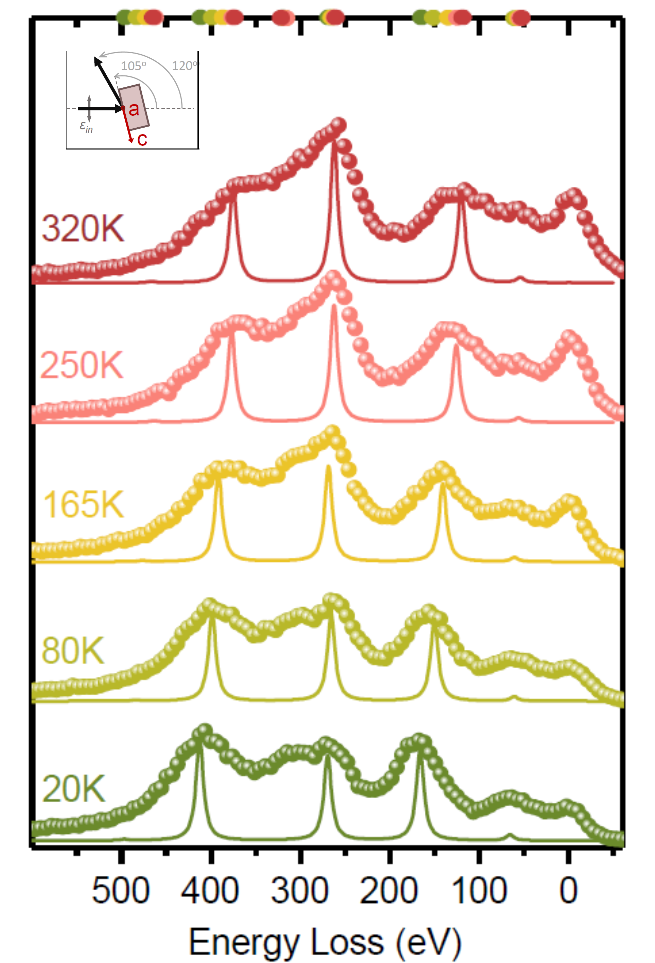}
    \caption{Temperature dependence of CeRh$_{3}$B$_2$ experimental RIXS spectra as measured on ID32 with incident photon energy $h\nu = \texttt{E}_{cen}$ and geometries as in inset. The ID32 data are extracted with the single photon algorithm (see Methods). The circles at the top show the peak positions for the simulations at the various temperatures. The lines are RIXS calculations performed with the crystal-field parameters in Table\,III, which provide the best fit of the data. }
\label{ID32_T}
\end{figure}

\begin{figure}[]
    \centering
    \includegraphics[width=0.8\columnwidth]{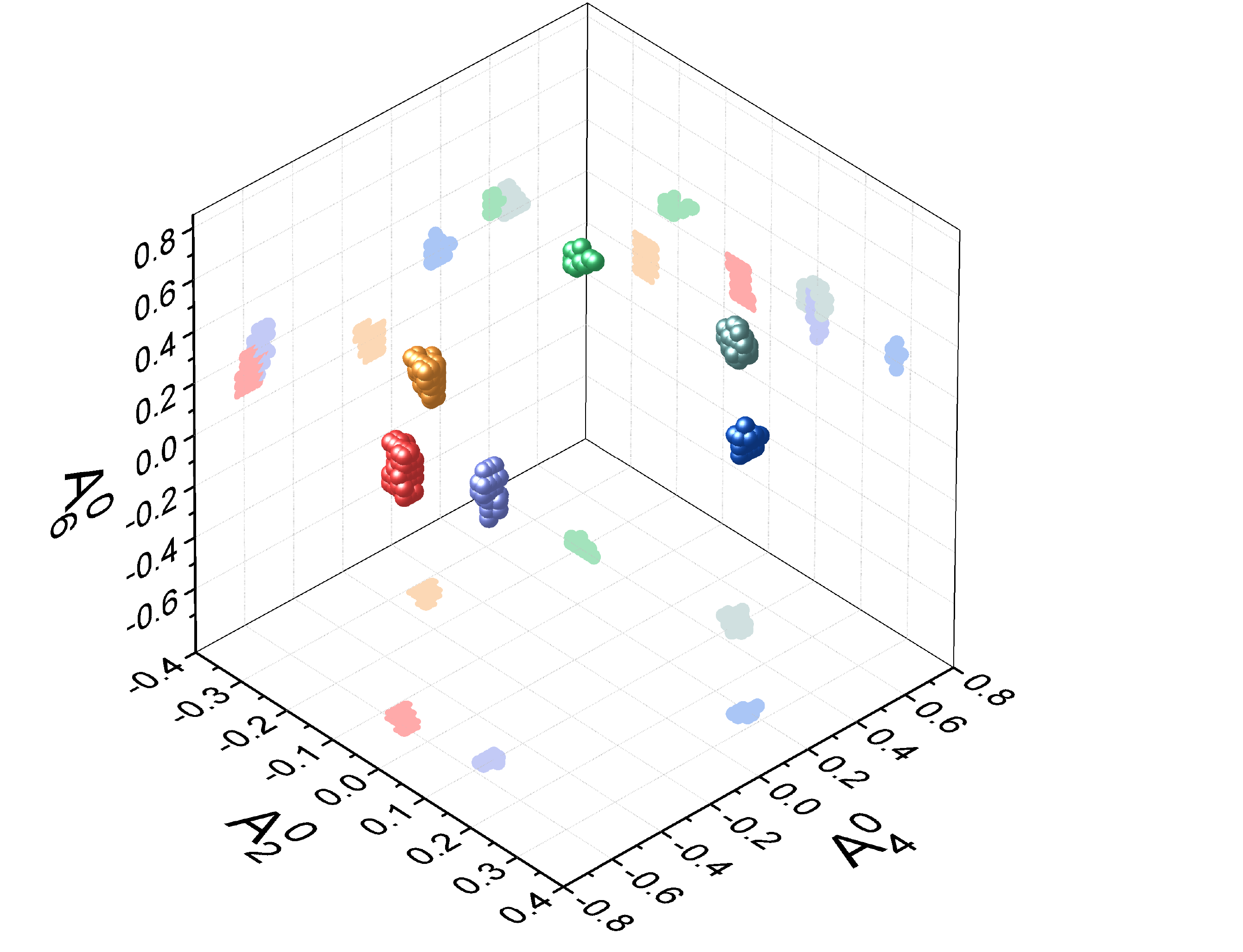}
    \caption{Regions of the parameter space yielding the crystal field splittings determined by the fit in Fig. \ref{RIXS}. The six different groups of crystal field parameter sets correspond each to a different arrangement of the three crystal-field levels of the lowest multiplet of Ce$^{3+}$.}
\label{parameters}
\end{figure}

Figure\,\ref{ID32_T} shows the RIXS spectra of CeRh$_3$Si$_2$ measured on ID32 at ESRF. Data were taken at 320, 250, 165, 80, and 20\,K. The data were taken with the geometry as in in Fig.\,\ref{RIXS} (left) and are extracted in the single photon alogorithm (see above) so that we can give a statistical error which is here about the size of the circles representing the data points. Also here the peaks at 130, 160, and 295\,meV move to higher energy transfers upon cooling, the other energies show only a minor temperature dependence. The colored dots at the top of Fig.\,\ref{ID32_T} indicate the peak positions at the various temperatures between 320\,K and 20\,K, as estimated by Voigt fits.  There is no sudden change of energies at T$_C$.

The lines in Fig.\,\ref{ID32_T} are the result of crystal-field calculations (see text above) with the parameters in Table\,IV.

\subsection{Finding crystal-field parameters and solutions}
\label{app:fit}
The $ A_2^0 $, $ A_4^0 $, and $ A_6^0 $ were scanned with a $41 \times 31 \times 31$ mesh in the region shown in Fig.\,\ref{parameters}. The $ A_6^6 $ parameter was found to have a minor effect on the crystal-field energies\,\cite{Givord2007INS}, and was scanned in steps of 0.15\,eV between $-0.6$\,eV and $0.6$. The spheres in Fig. \ref{parameters} represent the solutions compatible with the measured splittings (allowing for a $\pm 10$\,meV error to ensure that all possible solutions with similar splittings are found, despite the finite grid spacing used to span the parameter space).

\begin{figure*}[]
    \centering
     \includegraphics[width=0.86\columnwidth]{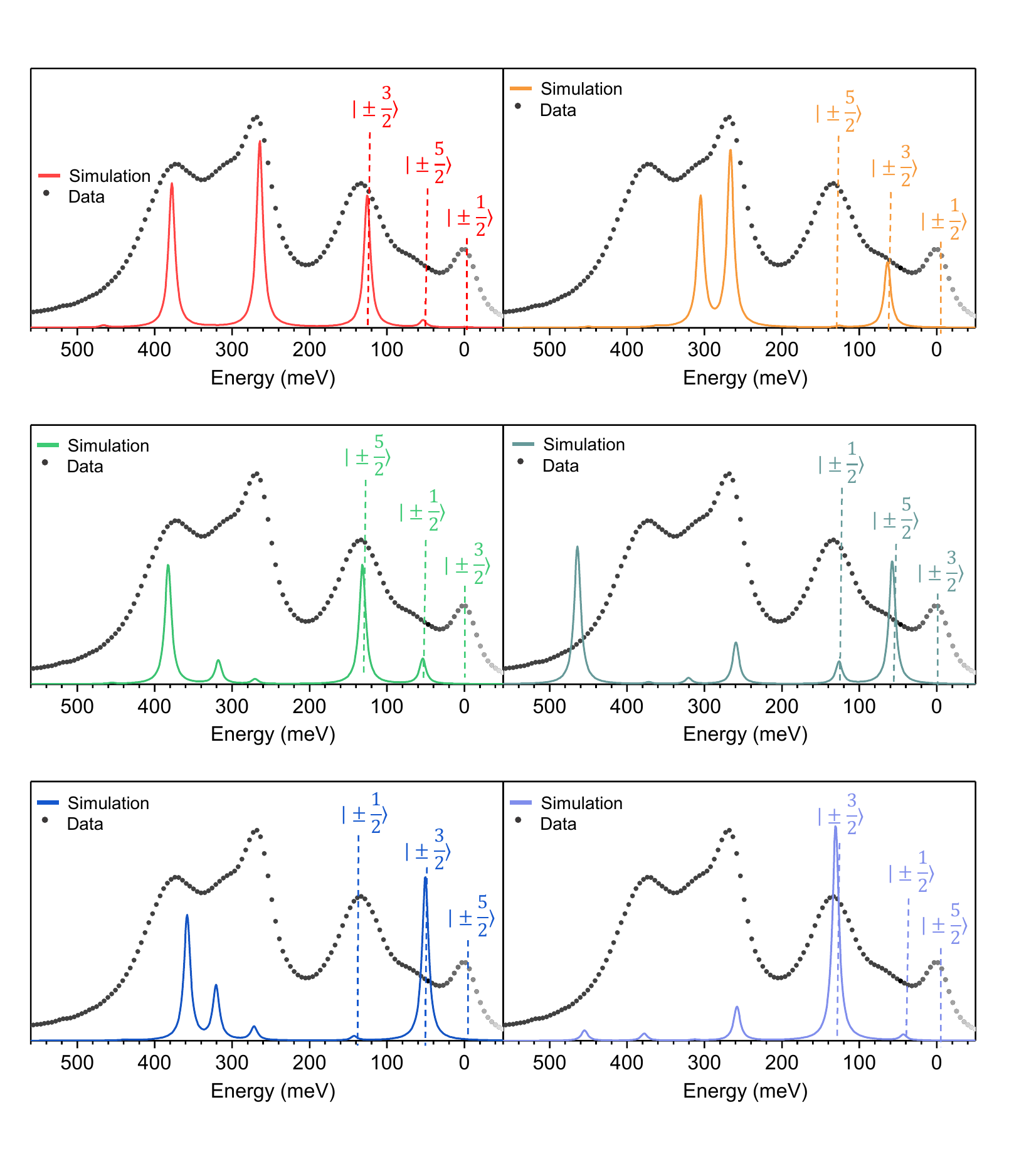}
     \includegraphics[width=0.86\columnwidth]{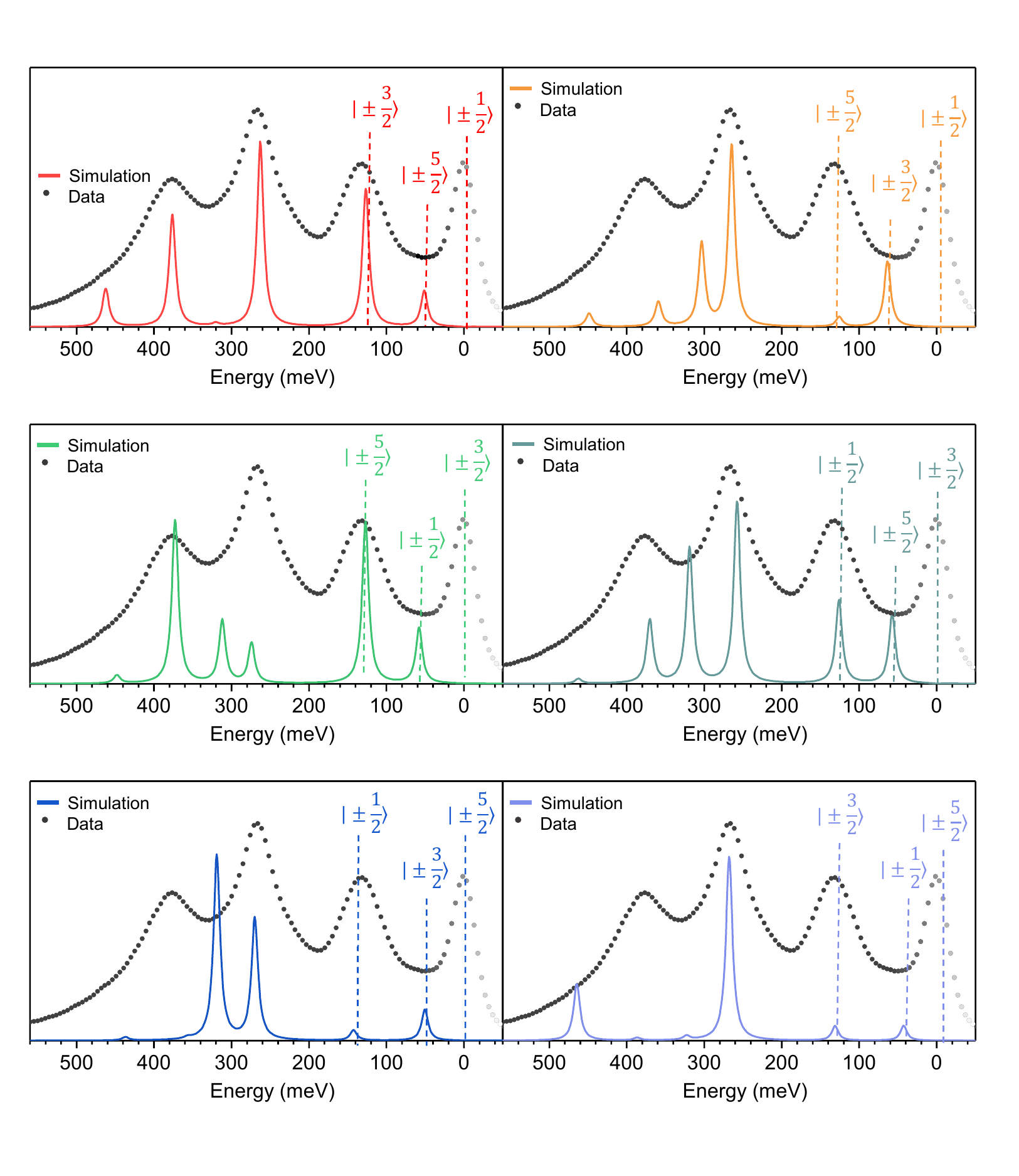}   
    \caption{RIXS calculations corresponding to the experimental spectra in Fig.\,\ref{RIXS} with E$_{in}$\,=\,E$_{cen}$, and $\phi$\,=\,105$^{\circ}$ (left) and 90$^{\circ}$ (right), respectively, using the six different sets of crystal-field parameters shown in Fig.\,\ref{parameters}. For each calculation, the main $\ket{\pm J_z}$ character of the first three crystal-field levels is indicated.}
\label{solutions}
\end{figure*}

The solutions are grouped in 6 regions of the phase space, which correspond to the 6 possible arrangements in energy of the three $\ket{J_z=\pm\frac{1}{2}}$, $\ket{J_z=\pm\frac{5}{2}}$ and $\ket{J_z=\pm\frac{7}{2}}$ doublets of the lowest $J=\frac{5}{2}$ multiplet. The corresponding orders of the first three $\ket{J_z}$ states are indicated in Fig.\,\ref{solutions} where we present the RIXS calculations for the 6 different sets of crystal-field parameters and compare them with the I21 experimental data of Fig.\,\ref{RIXS} for h$\nu$\,=\,E$_{cen}$. The sets of $A_k^m$ parameters correspond to the centers of gravity of each of the 6 regions in Fig.\,\ref{parameters}. The simulations are shown only for the inelastic region because the elastic intensity in the experimental spectra may strongly depend on the surface roughness and cannot be reliably calculated. We have excluded all solutions with the exception of the $red$ solution (top left, respectively) based on knowledge of previous findings: the ground state has to have $\pm J_z$\,=\,$\pm \frac{1}{2}$ symmetry and the state at about 130\,meV has to be a $\pm J_z$\,=\,$\pm \frac{3}{2}$ state to be observable with inelastic neutron scattering\,\cite{Givord2007INS}. The $red$ simulation fits the data reasonably well (see text above). 

\subsection{Lattice data}
\label{app:lattice}
\begin{figure}[H]
    \centering
    \includegraphics[width=0.7\columnwidth]{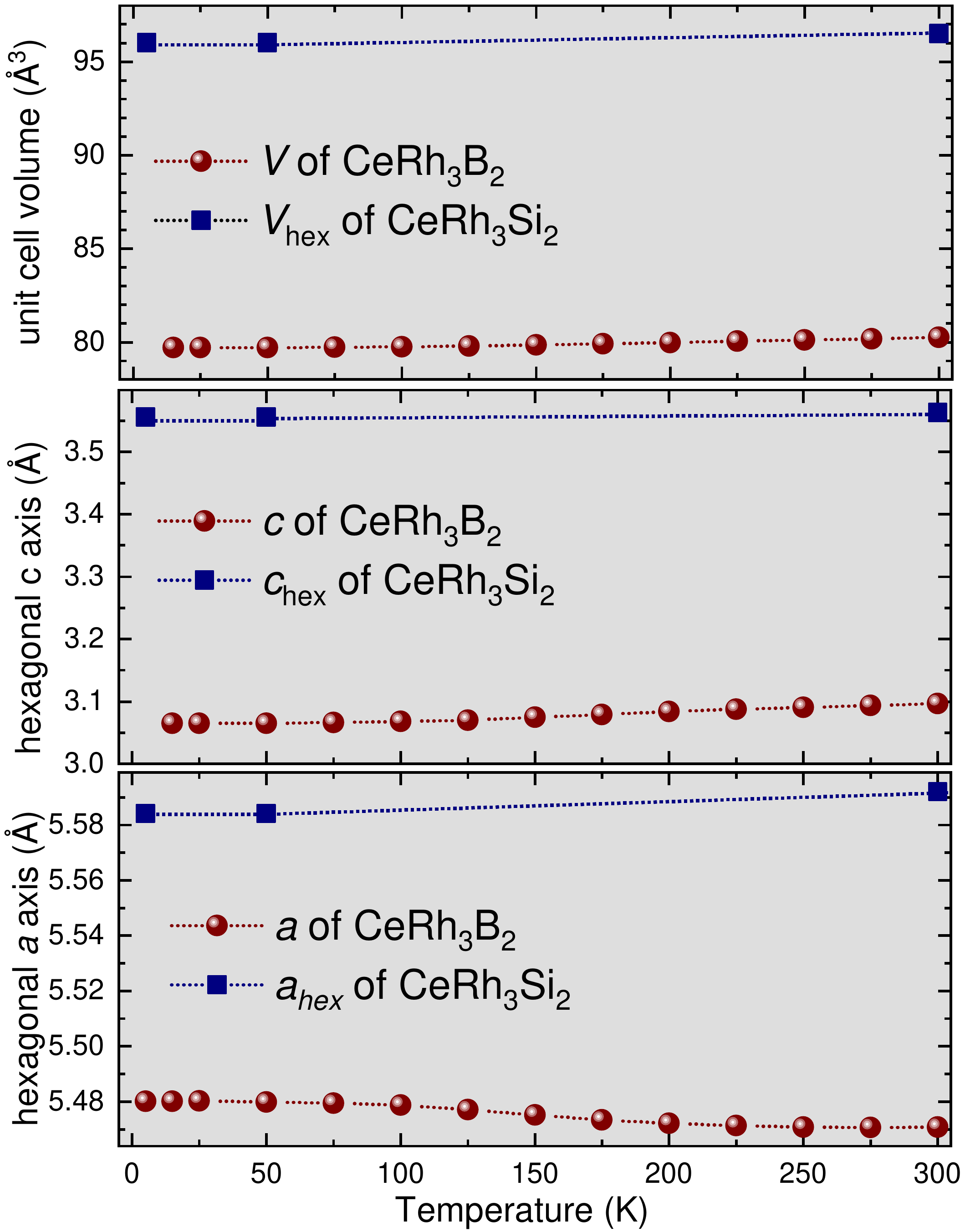}
    \caption{Lattice parameters of CeRh$_3$B$_2$ and of pseudo-hexagonal CeRh$_3$Si$_2$ from low $T$ to room temperature, (top) unit cell volume, (middle) hexagonal $c$ parameter, and (bottom) hexagonal $a$ parameter. The data of the boron compound are adapted from Ref.\,\cite{Langen1987} and of the silicon compound are calculated from the orthorhombic values given in\,\cite{Amorese2022}.}
\label{lattice}
\end{figure}
In CeRh$_3$Si$_2$ the orthorhombic distortion is small so that it can be considered as pseudo-hexagonal when using the orthorhombic $a$ axis as the hexagonal $c$-axis ($c_{hex}$)\,\cite{Amorese2022}. The comparison of the (pseudo)hexagonal lattice parameters of CeRh$_3$B$_2$ and CeRh$_3$Si$_2$ as function of temperature shows that the unit cell expands upon replacement of B by Si mainly due to expansion along the hexagonal $c$ axis i.e. due to the larger Ce--Ce distances in the Si compound (see Fig.\,\ref{lattice}). 

\subsection{Crystal-field parameters and energies at all temperatures}
\label{app:CEF}
Table\,III and IV list the crystal-field parameters and energies for all temperatures that correspond to the solution discussed in the main text. In Table\,IV the wave functions are labeled according to their majority $J_z$ contribution. 

\begin{table}[H]
\def\arraystretch{1.7}
\caption{Crystal field parameters (in meV) and spin-orbit interaction rescaling used for the RIXS simulations best fitting the data.}
\begin{tabular}{c|p{0.95cm}|p{0.95cm}|p{0.95cm}|p{0.95cm}|p{0.95cm}|p{0.95cm}}
\hline 
\rule[-1ex]{0pt}{2.5ex} Parameters          & \multicolumn{5}{c|}{ID32       }                     & I21 \\ 
\rule[-1ex]{0pt}{2.5ex}                     & 318\,K   & 250\,K   & 165\,K     & 80\,K   & 20\,K   & 295\,K  \\ 
\hline 
\rule[-1ex]{0pt}{2.5ex}  $ A_2^0 $    & -85      & -76      & -73        & -70     &  -69    & -80  \\
\hline                                                                                                      
\rule[-1ex]{0pt}{2.5ex}  $ A_4^0 $    & -600     & -605     & -640       & -685    &  -733   & -600 \\ 
\hline                                                                                                      
\rule[-1ex]{0pt}{2.5ex}  $ A_6^0 $    &  158     &  112      &  59        & 20      &  -13    &  113 \\ 
\hline                                                                                                      
\rule[-1ex]{0pt}{2.5ex}  $ A_6^6 $    & 150      & 110      & 110        & 110     &   120   & 150  \\
\hline                                                                                                     
\rule[-1ex]{0pt}{2.5ex}  $\zeta_{SO}$ (\%)  & 87       & 87       & 87.5       & 86      &   85    &  87   \\ 
\hline                                                
\end{tabular} 
\end{table}

\begin{table}[H]
\def\arraystretch{1.7}
\caption{Crystal-field scheme and splitting energies (in meV) resulting from the simulations best fitting the data.}
\begin{tabular}{c|c|c|c|c|c|c}
\hline 
\rule[-1ex]{0pt}{2.5ex}     Crystal field states              & \multicolumn{5}{c|}{ID32       }                     & I21 \\ 
\rule[-1ex]{0pt}{2.5ex}     $\ket{J,J_z}$                       								& 318\,K   & 250\,K   & 165\,K     & 80\,K   & 20\,K   & 295\,K  \\ 
\hline 
\rule[-1ex]{0pt}{2.5ex} $\ket{\frac{5}{2},\pm\frac{1}{2}}            								 	$   & 0        & 0        &  0     &  0      &    0    &    0  \\
\hline                                                                                                                      
\rule[-1ex]{0pt}{2.5ex} $\ket{\frac{5}{2},\pm\frac{5}{2}, \left(\mp\frac{7}{2}\right)}             		$   & 54       & 56       &  58    & 62      &   66    &   54  \\ 
\hline                                                                                                                                                     
\rule[-1ex]{0pt}{2.5ex} $\ket{\frac{5}{2},\pm\frac{3}{2}}             									$   & 120      & 126      & 136    & 153     &  166    &  125  \\                                        
\hline                                                                               
\rule[-1ex]{0pt}{2.5ex} $\ket{\frac{7}{2},\pm\frac{1}{2}}             									$   & 262      & 263      & 266    & 267     &  270    &  263  \\
\hline                                                                             
\rule[-1ex]{0pt}{2.5ex} $\ket{\frac{7}{2},\pm \frac{7}{2}}								$	& 323      & 317      & 319    & 318     &  321    &  323  \\ 
\hline                                                                                                                                                     
\rule[-1ex]{0pt}{2.5ex} $\ket{\frac{7}{2},\pm\frac{3}{2}}             									$   & 375      & 377      & 388    & 400     &  413    &  376  \\                                        
\hline                                                                             
\rule[-1ex]{0pt}{2.5ex} $\ket{\frac{7}{2},\mp\frac{5}{2}}								$	& 466      & 464      & 474    & 484     &  497    &  464  \\ 
\hline                                                
\end{tabular} 
\end{table}

\subsection{\textit{Ab-initio} derivation of parameters for the periodic Anderson impurity model}
\label{app:pAM}

The periodic Anderson impurity model (pAIM) contains the following terms:
\begin{equation}
\label{eq:pAM}
\begin{split}
  \hat{H}_\text{pAIM}=&\sum_{\mathbf{k}} \varepsilon_\mathbf{k}\hat{n}^b_\mathbf{k} + \sum_{i} \varepsilon_{J_z} \;\hat{n}^{J_z}_i+\sum_{i}  U\;\hat{n}^{J_z}_{\uparrow,i}\hat{n}^{J_z}_{\downarrow,i}+\\  &\sum_{i,\mathbf{k}} V_{J_z,c}(\mathbf{k})(f^\dagger_{i} c_\mathbf{k} + h.c.) +\mu \hat{N}^{tot}
\end{split}
\end{equation}
Here, $\varepsilon_\mathbf{k}$ is the dispersion relation of all uncorrelated conduction electrons, $\varepsilon_f$ stands for the on-site energy of the impurity levels (including local crystal electric field potentials), $U_i$ is the impurity on-site interaction, $V_{i,\mathbf{k}}$ is the coupling of impurity- and conduction orbitals and $\mu$ is the chemical potential. $\hat{n}$ are fermionic density operators, $\hat{N^{tot}}=\hat{n}^b+\hat{n}^f$ is the total electron density, and $f^\dagger$ and $f$ ($c^\dagger$ and $c$) are fermionic creation and annihilation operators of impurity electrons (conduction electrons). The sums over $i$ run over all impurity sites, the sums over $\mathbf{k}$ run over all lattice momenta $\mathbf{k}$ in the Brillouin zone. 

The corresponding single particle Green function is given by:
\begin{widetext}
	\begin{equation}
	\label{eq:GFpAM}
	G_\text{full}^\text{pAIM}(\omega,\mathbf{k})= \left((\omega+i\delta+\mu)\mathds{1}-
	\begin{pmatrix}
	\varepsilon_{J_z} & V_{J_z, c} (\mathbf{k}) \\
	V^*_{J_z,c}(\mathbf{k}) & \varepsilon_{c,\mathbf{k}} 
	\end{pmatrix}
	-
	\begin{pmatrix}
	\Sigma_{J_z}(\omega,\mathbf{k}) & 0 \\
	0 & 0
	\end{pmatrix}
	\right)^{-1}
	\end{equation}
\end{widetext}

with an implied limit of $\delta \rightarrow 0$. In the Green function, $\mu$ is the chemical potential and $\Sigma_{J_z}(\omega,\mathbf{k})$ is the self-energy for the $4f$-electrons which originates from the on-site interaction $U$ on the impurity sites. All quantities except for the self-energy can be extracted from the DFT calculation. Evaluating the impurity part of the Green function~\eqref{eq:GFpAM} leads to:
\begin{equation}
\label{eq:GFimp}
\begin{split}
G_{J_z}^\text{imp.}&(\omega,\mathbf{k})\equiv\left[G^\text{pAM}(\omega,\mathbf{k})\right]_{J_z,J_z}\\ =& \left((\omega+i\delta+\mu) - \varepsilon_{J_z} -\Sigma_{J_z}(\omega,\mathbf{k}) - \Delta_{J_z}(\omega,\mathbf{k})\right)^{-1}
\end{split}
\end{equation}
where $ \Delta_{J_z}(\omega,\mathbf{k})$ is the so called hybridization function
\begin{equation}
\label{eq:Delta}
\begin{split}
\Delta_{Jz}(\omega,\mathbf{k})&\equiv\sum_c \frac{|V_{J_z,c}(\mathbf{k})|^2}{\omega+i\delta+\mu-\varepsilon_{c,\mathbf{k}}} \\ &= \sum_c |V_{J_z,c}(\mathbf{k})|^2 \cdot G^0_c(\omega,\mathbf{k}) .
\end{split}
\end{equation}
In the second line we have the Green function of the uncorrelated conduction electrons before the hybridization with the Ce $4f$ states: $G^0_c(\omega,\mathbf{k}) \equiv (\omega+i\delta+\mu-\varepsilon_{c,\mathbf{k}})^{-1}$.

A closer look reveals a deep connection between the imaginary part of the hybridization function and the Kondo temperature scale as $|V_{J_z,c}(\mathbf{k})|^2 \propto {\cal J}$ is the effective spin coupling in the corresponding Kondo lattice model\,\cite{Sinjukow2002} and $-1/\pi\text{Im}[G^0_{c}(\omega,\mathbf{k})]=\rho^0_c(\omega,\mathbf{k})$ is the momentum resolved DOS of the conduction electrons. Their product determines the scale of $T_\text{K}$\,$\propto$\,$e^{-1/({\cal J}\rho^0_c)}$.

\section{acknowledgement} All authors thank A.C. Lawson for fruitful discussion and acknowledge support from the Max Planck-POSTECH-Hsinchu Center for Complex Phase Materials, the Diamond Light Source for RIXS beamtime on I21 under Proposal [SP18447], and the European Synchrotron Radiation Facility (ESRF) for RIXS beamtime on ID32 under proposal [HC3582]. A.S and A.A. benefited from Support of the German Research Foundation (DFG), Project No. 387555779. 

%


\end{document}